\documentclass[11pt,letter]{article}
\usepackage{amsmath,amsthm,latexsym,amssymb,amsfonts,epsfig}

\def\be{\begin{equation}}
\def\ee{\end{equation}}
\def\ba{\begin{eqnarray}}
\def\ea{\end{eqnarray}}

\newcommand\nn{\nonumber}
\newcommand\q{\quad}
\def\Nl{{\mathchoice
{\setbox0=\hbox{$\displaystyle\rm N$}\hbox{\hbox to0pt
{\kern0.4\wd0\vrule height0.9\ht0\hss}\box0}}
{\setbox0=\hbox{$\textstyle\rm N$}\hbox{\hbox to0pt
{\kern0.4\wd0\vrule height0.9\ht0\hss}\box0}}
{\setbox0=\hbox{$\scriptstyle\rm N$}\hbox{\hbox to0pt
{\kern0.4\wd0\vrule height0.9\ht0\hss}\box0}}
{\setbox0=\hbox{$\scriptscriptstyle\rm N$}\hbox{\hbox to0pt
{\kern0.4\wd0\vrule height0.9\ht0\hss}\box0}}}}
\def\Zl{{\mathchoice
{\setbox0=\hbox{$\displaystyle\rm Z$}\hbox{\hbox to0pt
{\kern0.4\wd0\vrule height0.9\ht0\hss}\box0}}
{\setbox0=\hbox{$\textstyle\rm Z$}\hbox{\hbox to0pt
{\kern0.4\wd0\vrule height0.9\ht0\hss}\box0}}
{\setbox0=\hbox{$\scriptstyle\rm Z$}\hbox{\hbox to0pt
{\kern0.4\wd0\vrule height0.9\ht0\hss}\box0}}
{\setbox0=\hbox{$\scriptscriptstyle\rm Z$}\hbox{\hbox to0pt
{\kern0.4\wd0\vrule height0.9\ht0\hss}\box0}}}}
\def\Ql{{\mathchoice
{\setbox0=\hbox{$\displaystyle\rm Q$}\hbox{\hbox to0pt
{\kern0.4\wd0\vrule height0.9\ht0\hss}\box0}}
{\setbox0=\hbox{$\textstyle\rm Q$}\hbox{\hbox to0pt
{\kern0.4\wd0\vrule height0.9\ht0\hss}\box0}}
{\setbox0=\hbox{$\scriptstyle\rm Q$}\hbox{\hbox to0pt
{\kern0.4\wd0\vrule height0.9\ht0\hss}\box0}}
{\setbox0=\hbox{$\scriptscriptstyle\rm Q$}\hbox{\hbox to0pt
{\kern0.4\wd0\vrule height0.9\ht0\hss}\box0}}}}
\def\Rl{{\mathchoice
{\setbox0=\hbox{$\displaystyle\rm R$}\hbox{\hbox to0pt
{\kern0.4\wd0\vrule height0.9\ht0\hss}\box0}}
{\setbox0=\hbox{$\textstyle\rm R$}\hbox{\hbox to0pt
{\kern0.4\wd0\vrule height0.9\ht0\hss}\box0}}
{\setbox0=\hbox{$\scriptstyle\rm R$}\hbox{\hbox to0pt
{\kern0.4\wd0\vrule height0.9\ht0\hss}\box0}}
{\setbox0=\hbox{$\scriptscriptstyle\rm R$}\hbox{\hbox to0pt
{\kern0.4\wd0\vrule height0.9\ht0\hss}\box0}}}}
\def\Cl{{\mathchoice
{\setbox0=\hbox{$\displaystyle\rm C$}\hbox{\hbox to0pt
{\kern0.4\wd0\vrule height0.9\ht0\hss}\box0}}
{\setbox0=\hbox{$\textstyle\rm C$}\hbox{\hbox to0pt
{\kern0.4\wd0\vrule height0.9\ht0\hss}\box0}}
{\setbox0=\hbox{$\scriptstyle\rm C$}\hbox{\hbox to0pt
{\kern0.4\wd0\vrule height0.9\ht0\hss}\box0}}
{\setbox0=\hbox{$\scriptscriptstyle\rm C$}\hbox{\hbox to0pt
{\kern0.4\wd0\vrule height0.9\ht0\hss}\box0}}}}
\def\Hl{{\mathchoice
{\setbox0=\hbox{$\displaystyle\rm H$}\hbox{\hbox to0pt
{\kern0.4\wd0\vrule height0.9\ht0\hss}\box0}}
{\setbox0=\hbox{$\textstyle\rm H$}\hbox{\hbox to0pt
{\kern0.4\wd0\vrule height0.9\ht0\hss}\box0}}
{\setbox0=\hbox{$\scriptstyle\rm H$}\hbox{\hbox to0pt
{\kern0.4\wd0\vrule height0.9\ht0\hss}\box0}}
{\setbox0=\hbox{$\scriptscriptstyle\rm H$}\hbox{\hbox to0pt
{\kern0.4\wd0\vrule height0.9\ht0\hss}\box0}}}}
\def\Ol{{\mathchoice
{\setbox0=\hbox{$\displaystyle\rm O$}\hbox{\hbox to0pt
{\kern0.4\wd0\vrule height0.9\ht0\hss}\box0}}
{\setbox0=\hbox{$\textstyle\rm O$}\hbox{\hbox to0pt
{\kern0.4\wd0\vrule height0.9\ht0\hss}\box0}}
{\setbox0=\hbox{$\scriptstyle\rm O$}\hbox{\hbox to0pt
{\kern0.4\wd0\vrule height0.9\ht0\hss}\box0}}
{\setbox0=\hbox{$\scriptscriptstyle\rm O$}\hbox{\hbox to0pt
{\kern0.4\wd0\vrule height0.9\ht0\hss}\box0}}}}
\newcommand{\ca}{\mathcal A}

\newcommand{\cc}{\mathcal C}

\newcommand{\ci}{\mathcal I}
\newcommand{\cj}{\mathcal J}
\newcommand{\ck}{\mathcal K}

\newcommand{\cm}{\mathcal M}
\newcommand{\cn}{\mathcal N}

\newcommand{\cs}{\mathcal S}
\newcommand{\ct}{\mathcal T}

\newcommand{\cy}{\mathcal Y}

  \newcommand{\Fc}{\mathfrak{C}}

\newcommand{\td}{\text{d}}

\begin{document}

\title{Partial and Complete Observables for Canonical General Relativity}

\author{B. Dittrich\thanks{dittrich@aei.mpg.de and bdittrich@perimeterinstitute.ca}, \\
 MPI f. Gravitationsphysik, Albert-Einstein-Institut, \\
 Am M\"uhlenberg 1, 14476 Golm near Potsdam, Germany \\
and \\
Perimeter Institute for Theoretical Physics \\
31 Caroline Street North, Waterloo, ON N2L 2Y5, Canada \\
}

\date{{\small Preprint  AEI-2005-128}}

\maketitle

\begin{abstract}

In this work we will consider the concepts of partial and complete observables for canonical general relativity. These concepts provide a method to calculate Dirac observables. The central result of this work is that one can compute Dirac observables for general relativity by dealing with just one constraint. For this we have to introduce spatial diffeomorphism invariant Hamiltonian constraints. It will turn out that these can be made to be Abelian. Furthermore the methods outlined here provide a connection between observables in the space--time picture, i.e. quantities invariant under space--time diffeomorphisms, and Dirac observables in the canonical picture. 

\end{abstract}

\section{Introduction}

The most important feature of general relativity is that space--time itself is dynamical and hence that no reference frame is preferred. This causes a lot of difficulties if one wants to quantize the theory because in its non--relativistic formulation quantum theory requires a fixed background and a preferred splitting of space--time into space and time. Nevertheless a well--developed approach for quantizing general relativity is to apply canonical quantization to the theory, as is done for instance in Loop Quantum Gravity \cite{1.1,7.2,7.3}. However the fact that no reference frame is preferred in general relativity causes the appearence of constraints, which restrict the set of phase space points whith a physical interpretation. These constraints do also generate gauge transformations which in the canonical theory among other things translate between different splittings of the space--time into space and time.

If one wants to quantize a theory with gauge symmetries one has to look for physical observables, also called Dirac observables, i.e. phase space functions which are invariant under gauge transformations. For general relativity this is a very difficult problem since here also translations in time are gauge transformations. This means that one has to solve at least partially the dynamics of general relativity in order to obtain gauge invariant quantities. Because this dynamics is described by a complicated system of highly non--linear partial differential equations it is not surprising that there are almost no gauge invariant phase space functions known.\footnote{
For the case of gravity in four space--time dimensions and for asymptotically flat boundary conditions there are 10 gauge invariant phase space functions known. These are the ADM charges \cite{ADM} given by the generators of the Poincare transformations at spatial infinty. Additionally an observable is known, which takes only a few discrete values and is trivial on almost all points in phase space \cite{jacobson}. For gravity coupled to matter, in some cases gauge invariant functions describing matter are known but in general no phase space functions which describe the gravitational degrees of freedom (with the exception of the ADM charges). 
 Yet there are infinitely many gauge invariant degrees of freedom.
}

Therefore the hope is that one can at least develop an approximation scheme for Dirac observables. To this end it is valuable to have methods with which one can find Dirac observables in principle. One such method was proposed by Rovelli in \cite{RovPartObs} for systems with one gauge degree of freedom. Let us assume that this gauge degree of freedom corresponds to reparametrizations of an (unphysical) time parameter. Choose two {\it partial observables} $f$ and $T$, that is gauge variant phase space functions. The partial observable $T$ will serve as a clock, hence we will call it a {\it clock variable}. The {\it complete observable} $F_{[f;\,T]}(\tau)$ expresses the gauge invariant relation between $f$ and $T$: It gives the value of $f$ at that moment (i.e. at that unphysical time parameter) at which $T$ assumes the value $\tau$. Here $\tau$ can be chosen as any value from the range of the function $T$. 

Hence (for systems with one constraint) complete observables are one--parameter families of Dirac observables. One advantage of the concept of partial and complete observables is, that it has an immediate physical interpretation. This might help in order to develop an approximation scheme. Calculating complete observables is the same as making predictions and the precision with which we can do these predictions will depend on the dynamics of the theory and on the phase space region one is considering.   
 
Now general relativity is a system with infinitly many gauge degrees of freedom. Therefore \cite{bd1} generalized the concepts of partial and complete observables to systems with an arbitrary number of constraints. The main idea for that is to use instead of one clock variable $T$ as many clock variables as there are constraints. Furthermore it was shown that complete observables can be written as a (formal) power series. Among other things this was used in order to calculated the Poisson algebra of complete observables and also to show that it is possible to calculate complete observables in stages, i.e. first to compute (partially) complete observables with respect to a subalgebra of the constraints and then to use these partially complete observables for the calculation of complete observables with respect to all constraints. However in the second step one can ignore the subalgebra of the constraints which were used in the first step and just deal with the remaining constraints.

Ideas how to use complete observables in a quantization of a gauge system and more specifically general relativity can be found in \cite{RovPartObs},\cite{reducedphasespaceq} and in \cite{jb}.

In this work we will apply the methods developed in \cite{bd1} to general relativity. One central result of this work is a variation of the method to calculate complete observables in stages. Namely we will show that for a certain choice of partial observables it is possible to compute complete observables for general relativity by dealing with just {\it one constraint instead of infinitely many constraints}. 

But first we will give an introduction of the concepts of partial and complete observables and of the relevant results of \cite{bd1} in section \ref{completefields}. Right from start we will consider field theories, that is systems with infinitely many constraints. In section \ref{completefields} we will also explain under which circumstances one can reduce the number of constraints one has to deal with for the calculation of complete observables from infinitely many to finitely many constraints.

Section \ref{gr} reviews the canonical formulation of general relativity, in particular the connection between the canonical picture, i.e. where space--time is splittet into space and time and the covariant picture, where one considers space--time without such a splitting.

This is necessary to understand space--time scalars in the canonical picture. As will be explained in section \ref{completescalars} one can use these space--time scalars as partial observables in order to reduce the number of constraints one has to deal with for the computation of complete observables drastically. Moreover we will show how one can construct canonical fields which behave as space--time scalars in section \ref{spacetimemetric} and in appendix \ref{appendix1}.

In section \ref{solvediffeo} we explain how one can get rid off the spatial diffeomorphism constraints. That means that afterwards one works only with quantities, which are invariant under spatial diffeomorphisms. In particular we introduce spatial diffeomorphism invariant Hamiltonian constraints. Here we will show that using special partial observables one has just to deal with one constraint in order to calculate Dirac observables invariant under all constraints. 
 
Section \ref{AbelianConstraints} discusses the Poisson brackets of the spatial diffeomorphism invariant Hamiltonian constraints introduced in section \ref{solvediffeo} and shows how one can get Abelian diffeomorphism invariant Hamiltonian constraints. 

In section \ref{example} we will consider the example of gravity coupled to matter scalar fields and end with a summary in \ref{summary}.

\section{Complete Observables for Field Theories}\label{completefields}

In this section we will consider partial and complete observables for field theories and at the same time summarize those results from \cite{bd1} which are necessary for this work.

For a field theory we work with a phase space $\cm$ which is some Banach space of fields given on a spatial $d$--dimensional manifold $\Sigma$ and subject to some boundary conditions. Points in $\Sigma$ will be denoted by $\sigma$ and we will assume that coordinates $(\sigma^a)_{a=1}^d$ have been fixed. We will also denote the coordinate tuple $(\sigma^a)_{a=1}^d$ by $\sigma$. Phase space points in $\cm$ will be denoted by $x$ or $y$. 

The symplectic structure of this phase space is defined via canonical coordinates denoted by $(\phi^A(\sigma),\pi_A(\sigma); \sigma \in \Sigma)$ where $A$ is from some finite index set $\ca$. The non--vanishing Poisson brackets are given by 
\ba\label{fields1}
\{\phi^A(\sigma),\pi_B(\sigma')\}=\alpha_A \delta^A_{B}\,\delta(\sigma,\sigma')
\ea
where $\delta(\sigma,\sigma')$ is the delta function on $\Sigma$ and $\alpha_A$ is a coupling constant for the fields $\phi^A,\pi_A$.

Constraint field theories have an infinite set of constraints $C_K(\sigma)$, labelled by an index $K$ from some finite index set $\ci$ and by the points $\sigma$ of $\Sigma$. We will assume that all these constraints generate gauge transformations and are hence first class, that is, that the Poisson bracket of two constraints vanishes on the constraint hypersurface, i.e. that hypersurface in phase space, where all the constraints vanish. Furthermore we will assume that all gauge transformations can be generated by these constraints: I.e. consider the smeared constraint
\ba\label{july1}
C[{ \Lambda}]:=\int_\Sigma \Lambda^K(\sigma) C_K(\sigma) \, \td^d \sigma
\ea
where $\Lambda^K(\sigma)$ are phase space independent smearing function, $\td^d \sigma$ denotes the coordinate volume element on $\Sigma$ and we sum over the repeated index $K\in \ci$. The gauge transformation $\alpha^t_{C[{\Lambda}]}$ generated by this constraint acts on a field $\psi$ as 
\ba\label{july2}
\alpha^t_{C[{ \Lambda}]}(\psi(\sigma))=\sum_{r=0}^\infty \frac{t^r}{r!} \{\psi(\sigma), C[{ \Lambda}]\}_r \q .
\ea
Here $\{g,C\}_r$ are iterated Poisson brackets, i.e. $\{g,C\}_0=f$ and $\{g,C\}_{r+1}=\{\{g,C\}_r,C\}$. We will often set the parameter $t$ to $t=1$ and omit the corresponding index, gauge transformations for other values of $t$ can be obtained by rescaling the smearing functions $\Lambda^K$.

Complete observables are phase space functions, which are invariant under such gauge transformations, i.e. which are Dirac observables. Complete observables are associated to partial observables, these are phase space functions, which are in general not invariant under gauge transformations. These partial observables are divided into clock variables $T^K(\sigma),\, K\in \ci,\, \sigma\in \Sigma $ and another partial observable $f$. We need as many clock variables as there are constraints, therefore these are labelled by the index $K\in \ci$ and by the points $\sigma\in \Sigma$. In general the partial observables may be arbitrary functionals of the basic canonical fields, but in the following we will often choose the partial observables also as fields, i.e. as functionals which will give the value of some field, built from the basic canonical fields, at a certain point $\sigma_0\in \Sigma$. 

The complete observable $F_{[f;\,T]}(\tau,x)$ associated to the partial observable $f$ and the clock variables $T:=\{T^K(\sigma)\}_{K\in \ci, \sigma\in \Sigma}$ will in general depend on infinitely many parameters $\tau:=\{\tau^K(\sigma)\}_{K\in \ci, \sigma\in \Sigma}$. It gives the value of the phase space function $f$ at that point $y$ in the gauge orbit through $x$ at which the the clock variables give the values $\left[T^K(\sigma)\right](y)=\tau^K(\sigma)$ for all $K\in \ci$ and all $\sigma\in \Sigma$.

The conditions $\left[T^K(\sigma)\right](y)=\tau^K(\sigma)$ should specify a unique\footnote{This uniqueness is not always necessary, as is discussed in \cite{bd1}.}
% However a criterium, %which is necessary in order to obtain 
%for this uniqueness is that the functional determinant $\text{det}(\{T^K(\sigma),C_{K'}(\sigma')\})_{K,K'\in \ci;\,\sigma,\sigma'\in \Sigma}$ does not vanish.
%\footnote{The conditions under which complete observables are well--defined are discussed in \cite{bd1}.} 
point in the gauge orbit through $x$. At that point one evaluates the phase space function $f$ and this is the value the complete observable $F_{[f;\,T]}(\tau,x)$ assumes on all points on the gauge orbit through $x$. Hence complete observables are gauge invariant.  

One way to calculate the value of the complete observable $F_{[f;\,T]}(\tau,x)$ is the following: First find the point $y$ in the gauge orbit through $x$ at which $\left[T^K(\sigma)\right](y)=\tau^K(\sigma)$. That is, determine the flow
\ba\label{july3}
\left[\alpha_{C[\Lambda]}(T^K(\sigma))\right](x)=\sum_{r=0}^\infty \frac{1}{r!}\{T^K(\sigma),C[\Lambda]\}_r \,(x)
\ea
and find functions $\beta^K(\sigma)$ such that
\ba \label{july4}
\left[\alpha_{C[\Lambda]}(T^K(\sigma))\right]_{\Lambda \rightarrow \beta(x)}(x) \simeq \tau^K(\sigma)
\ea
for all $K\in \ci$ and $\sigma\in \Sigma$. Here the symbol $\simeq$ means, that the equation needs only to hold weakly, i.e. on the constraint hypersurface, or in other words, modulo terms which are at least linear in the constraints. In contrast to the smearing functions $\Lambda^K$, which we assumed to be phase space independent, the functions $\beta^K$ will in general depend on the phase space point $x$. Therefore in equation (\ref{july4}) one has first to calculate the gauge transformation for general phase space independent $\Lambda^L(\sigma)$ and afterwards to replace $\Lambda^L(\sigma)$  by the phase space dependent functions $\beta^L(\sigma)$. The value of the complete observable is then given by
\ba\label{july5}   
F_{[f;\,T]}(\tau,x)\simeq \left[\alpha_{C[\Lambda]}(f)\right]_{\Lambda \rightarrow \beta(x)} (x) \q .
\ea

In most cases, it will be very difficult to find an explicit expression for the flows (\ref{july3}) and to solve the equations (\ref{july4}). In \cite{bd1} we therefore developed a system of partial differential equations, which describe the complete observables. For field theories one will get functional differential equations. In order to state these functional differential equations, we remark, that one can always replace one constraint set by another set of constraints as long as these define the same constraint hypersurface. Both set of constraints will then lead to the same gauge orbits at least on the constraint hypersurface. The main idea is now, to introduce new constraints, which will evolve the clock variables in a particularly simple way:

Consider the infinte--dimensional matrix defined by
\ba\label{mj1}
{A^K}_{L}(\sigma,\sigma'):=\{ T^K(\sigma),C_L(\sigma')\} \q .
\ea
One has to find the inverse ${(A^{-1})^L}_{M}(\sigma,\sigma')$ to this matrix, i.e. an integral kernel ${(A^{-1})^L}_{M}(\sigma',\sigma'')$ satisfying
\ba\label{mj2}
\int_\Sigma \! \td^d \sigma' {A^K}_L(\sigma,\sigma'){(A^{-1})^L}_{M}(\sigma',\sigma'') &=& \delta^K_{M}\,\delta(\sigma,\sigma'') \nn \\
\int_\Sigma\! \td^d \sigma' {(A^{-1})^K}_{L}(\sigma,\sigma'){A^L}_{M}(\sigma',\sigma'') &=& \delta^K_{M}\,\delta(\sigma,\sigma'') \q .
\ea
%where $\td^d \sigma$ denotes the coordinate volume element on $\Sigma$. 
Then the new constraints $\tilde C_K$ are given by
\ba\label{mj3}
\tilde C_K(\sigma) :=\int_\Sigma \! \td^d \sigma' \, C_L(\sigma') {(A^{-1})^L}_{K}(\sigma',\sigma)  \q .
\ea
Note that the integral kernel $(A^{-1})^L_{M}(\sigma',\sigma'')$ is easier to obtain if $A_{KL}(\sigma,\sigma')$ is ultra--local, i.e. $A_{KL}(\sigma,\sigma') \sim \delta(\sigma,\sigma')$.

With these new constraints, we have that at least on the constraint hypersurface
\ba\label{july6}
\{T^K(\sigma), \tilde C_L(\sigma')\}\simeq \delta^K_L \,\delta(\sigma,\sigma')
\ea
and hence
\ba\label{july7}
\alpha_{\tilde C[\Lambda]}(T^K(\sigma))\simeq T^K(\sigma)+ \Lambda^K(\sigma)
\ea
where $\tilde C[\Lambda]:=\int_\Sigma \Lambda^L(\sigma)\,\tilde C_L(\sigma)\td^d \sigma$. 

Equations (\ref{july6},\,\ref{july7}) mean that the flow generated by the constraint $\tilde C_K(\sigma)$ is along the $T^K(\sigma)$--coordinate line (i.e. along the one--dimensional lines, where all the other clock variables and all Dirac observables are constant) on the constraint hypersurface. From this it follows also, that the flows generated by the new constraints have to commute on the constraint hypersurface, i.e. that the Poisson brackets between two of the new constraints
\ba\label{july8}
\{\tilde C_K(\sigma),\tilde C_L(\sigma')\}=O(C^2)
\ea
involve only terms, which are at least quadratic in the constraints.\footnote{Indeed, equation (\ref{july8}) can be checked explicitly, see \cite{bd1}.}
 We will therefore call these constraints weakly Abelian.  

With these new constraints solving the equations (\ref{july3},\,\ref{july4}) becomes trivial and we can plug in the solutions $\beta^L(\sigma)=\tau^L(\sigma)-T^L(\sigma)$ into equation (\ref{july5}) in order to obtain for the complete observable
\ba\label{july9}
F_{[f;\,T]}(\tau,x)
\,&\simeq& \, 
\left[\alpha_{\tilde C[\Lambda]}(f)\right]_{\Lambda\rightarrow (\tau-T(x))}(x) 
\nn\\
\,&\simeq& \,
\sum_{r=0}^\infty \,\frac{1}{r!}
\int_\Sigma \!\td^d \sigma_1 \cdots \td^d \sigma_r \,\,
\{\cdots\{f, \tilde C_{K_1}(\sigma_1)\},\cdots,\tilde C_{K_r}(\sigma_r)\}(x) \times 
\nn \\
&&\q\q\q\q\q \q\q
(\tau^{K_1}(\sigma_1)-T^{K_1}(\sigma_1)(x))\cdots
(\tau^{K_r}(\sigma_k)-T^{K_r}(\sigma_r)(x)) \q . \q
\ea

From here it is straightforward to see, that complete observables satisfy the functional differential equations
\ba\label{july10}
\frac{\delta}{\delta \,\tau^K(\sigma)}F_{[f;\,T]}(\tau,x) \simeq F_{[\{f,\tilde C_K(\sigma)\};\,T]}(\tau,x)
\ea
with initial conditions
\ba \label{july11}
F_{[f;\,T]}(\tau \equiv T ,\cdot)=f \q .
\ea
%\ba \label{july11}
%F_{[f;\,T]}(\tau,\cdot)_{| \{T^K\equiv \tau^K \,\forall\, K\in \ci\}} \simeq f_{|\{ T^K\equiv \tau^K \,\forall\, K\in \ci\}} \q .
%\ea

Note that equation (\ref{july9}) gives complete observables as a formal power series in the $(\tau^K(\sigma)-T^K(\sigma))(x)$.
Many properties of complete observables can be proven by using this power series, see \cite{bd1}.%e.g. Poisson brackets between complete observables

For instance consider the case, that we have clock variables $T^K(\sigma)$ and a partial observable $f=\psi(\sigma^*)$ such that these fields have at least weakly ultra--local Poisson brackets with the constraints, i.e. such that 
\ba\label{july13}
\{T^K(\sigma), C_L(\sigma')\} \sim \delta(\sigma,\sigma') \q\text{and} \q \{\psi(\sigma^*),C_L(\sigma')\}\sim \delta(\sigma^*,\sigma')
\ea
modulo terms at least linear in the constraints. Then the matrices ${A^K}_L(\sigma,\sigma')$ and ${(A^{-1})^K}_L(\sigma,\sigma')$ will be also ultra--local\footnote{Terms which are not ultra--local and hence are vanishing on the constraint hypersurface can be omitted.} and the new constraints are just given by
\ba\label{july14}
\tilde C_L(\sigma)={(B^{-1})^K}_L(\sigma)C_K(\sigma)
\ea
where the phase space functions ${B^K}_L(\sigma)$ are defined by 
\ba\label{july15}
{A^K}_L(\sigma,\sigma')=:{B^K}_L(\sigma)\delta(\sigma,\sigma')\q .
\ea 

Now we have also for the new constraints that
\ba\label{july15a}
\{\psi(\sigma^*),\tilde C_L(\sigma)\}\sim \delta(\sigma^*,\sigma)
\ea
at least on the constraint hypersurface. Hence the first order term in the power series (\ref{july11}) for the complete observable $F_{[\psi(\sigma^*),T]}(\tau,\cdot)$ can be written as
\ba\label{july16}
 \int  \{\psi(\sigma^*),\tilde C_K (\sigma)\} (\tau^K(\sigma)-T^K(\sigma)) \,\td^d \sigma                               &\simeq&\! \int \{\psi(\sigma^*), \tilde C_K[1]\} \delta(\sigma^*,\sigma)\,(\tau^K(\sigma)-T^K(\sigma)) 
\,\td^d \sigma
\nn\\
&\simeq&\!
\{\psi(\sigma^*), \tilde C_K[1]\} (\tau^K(\sigma^*)-T^K(\sigma^*)) \q .\nn\\
\ea
Here $\tilde C_K[1]$ are the smeared constraints $\tilde C_K[\Lambda_{(K)}]:=\int \Lambda_{(K)}(\sigma) \tilde C_K(\sigma)\td^d \sigma$ with some smearing functions $\Lambda_{(K)}$ such that $\Lambda_{(K)}(\sigma^*)=1$. Note that this first order term just depends on the parameters $\tau^K(\sigma^*), K\in \ci$ and not on the parameter fields $\tau^K(\sigma),\,K\in \ci,\sigma\in \Sigma$. The question is whether this holds also for the higher order terms. Consider the double Poisson bracket of $\psi(\sigma^*)$ with two constraints:
\ba\label{july17}
\{\{\psi(\sigma^*),\tilde C_{K_1}(\sigma_1)\},\tilde C_{K_2}(\sigma_2)\} & \simeq& \{\psi(\sigma^*),\tilde C_{K_1}[1]\}, \tilde C_{K_2}(\sigma_2)\} \,\delta(\sigma^*,\sigma_1) \nn\\
&\simeq&  \{\psi(\sigma^*),\tilde C_{K_2}(\sigma_2)\},\tilde C_{K_1}[1]\}\,\delta(\sigma^*,\sigma_1) \nn\\
&\simeq& \{\psi(\sigma^*),\tilde C_{K_1}[1]\},\tilde C_{K_2}[1]\}\,\delta(\sigma^*,\sigma_1)\delta(\sigma^*,\sigma_2)\q  \q\q\q
\ea
where we used in the second and third line that the new constraints are weakly Abelian.
Iterating this argument we will find
\ba\label{july18}
&&\{\cdots\{\psi(\sigma^*),\tilde C_{K_1}(\sigma_1)\},\cdots, \tilde C_{K_r}(\sigma_r)\}
\simeq  \nn\\
&&\q\q\q\q\q\q\q\q\q\q
\{\cdots\{\psi(\sigma^*),\tilde C_{K_1}[1]\},\cdots, \tilde C_{K_r}[1]\}\, \delta(\sigma^*, \sigma_1)\cdots \delta(\sigma^*,\sigma_r) \q. \q\q\q \q\q
\ea
 Now the power series (\ref{july9}) for the complete observable $F_{[\psi(\sigma^*),T]}(\tau,\cdot)$ simplifies very much
since all integrals over the spatial manifold $\Sigma$ are trivially solved by delta functions:
\ba\label{fields21}
F_{[\psi(\sigma^*);T]}(\tau,\cdot)
&\simeq&
\sum_{r=0}^\infty \frac{1}{r!}\{\cdots\{\psi(\sigma^*),\tilde C_{K_1}[1]\},\cdots, \tilde C_{K_r}[1]\} 
\nn\\
 &&\q\q\q\q
(\tau^{K_1}(\sigma^*)-T^{K_1}(\sigma^*))\cdots(\tau^{K_r}(\sigma^*)-T^{K_r}(\sigma^*)) \q . \q\q
\ea

In this power series we have just to deal with the {\it finitely many constraints} $\tilde C_K[1],\,K\in\ci$ and not with the infintely many constraints $\tilde C_K(\sigma),\,K\in\ci,\,\sigma\in\Sigma$ as is the case in the general formula (\ref{july9}). Also the complete observable just depends on the finitely many parameters $\tau^K(\sigma^*)$ and not on the whole parameter fields $\tau^K(\sigma),\sigma\in\Sigma$. In correspondence with this we are just left with the clock varibales $T^K(\sigma^*)$ at the point $\sigma^*$. 

We arrived at this reduction by utilising two facts: Firstly we used partial observables which are already invariant with respect to almost all constraints, i.e. with respect to all constraints except those at the point $\sigma^*$. In \cite{bd1} it is outlined that such 'partially invariant' partially observables can be used, in order to construct complete observables with respect to all constraints.

Secondly we used weakly Abelian constraints. This ensured that the partial observables remain `partially invariant' if one applies the Poisson brackets with these constraints. 

In the next section we will review the canonical formulation of general relativity and in the following section ask for phase space functios which have ultra--local Poisson brackets with the constraints.

\section{Canonical Formulation of General Relativity} \label{gr}

%Solving the equations of motions means to determine the canonical fields on a sequence of spatial hypersurfaces $\Sigma_t$ labelled by a time-parameter $t$. Taking all the fields on this sequence of hypersurfaces together one will restore the geometry of the space--time manifold $\cs$. Moreover the embeddings of the hypersurfaces $\Sigma_t$ into the space--time manifold define a foliation of space--time. So a solutions of the canonical equations of motions does not only give us the geometry of the space--time manifold $\cs$ but also a foliation of this space--time manifold into spatial hypersurfaces $\Sigma_t$.

For extended reviews of the canonical formulation of general relativity see \cite{ADM,7.3,hyperspace,wald}. We will here just summarize the relevant results.

The phase space variables for the gravitational degrees of freedom are given by the spatial metric $g_{ab}(\sigma)$ and the conjugated momentum $p^{ab}(\sigma)$, which are fields on the spatial manifold $\Sigma$.\footnote{We will consider here the canonical formulation in the ADM-variables \cite{ADM}.} We will assume that $\Sigma$ is a compact manifold. The Poisson brackets between these fields are given by
\ba
\{g_{ab}(\sigma),p^{cd}(\sigma')\}=\kappa \,\delta^c_{(a}\delta^d_{b)}\delta(\sigma,\sigma')
\ea
where $\kappa=8\pi G/c^3$ with $G$ Newton's constant is the graviational coupling constant.
The matter degrees of freedom are similarly described by matter fields defined on $\Sigma$. We will consider only non--derivative couplings of matter to gravity.    

 The phase space variables are subject to constraints. For each point $\sigma \in \Sigma$ there are $d$ diffeomorphismen constraints 
\ba\label{fields3}
 C_a(\sigma)={}^{gr}C_a(\sigma)+{}^{mat}C_a(\sigma)=-\frac{2}{\kappa}\,g_{ac}D_b p^{bc}(\sigma)+{}^{mat}C_a(\sigma)
\ea
where $D$ denotes the covariant (Levi--Civita) differential associated to the metric $g_{ab}$. The matter parts of the diffeomorphism constraints depends on the tensorial character of the matter fields. If these are described by spatial scalar fields $\varphi^K$ and conjugated momenta $\pi_K$, which have to be scalar densities, the matter parts are given by
\ba
{}^{mat}C_a(\sigma)=\sum_K \frac{1}{\alpha_K}\varphi^A_{,a}\pi^A \q \q .
\ea
The subindex `$\,{,a}\,$' denotes partial differentiation with respect to $\sigma^a$.
Furthermore we have the so called Hamiltonian constraints
\ba\label{fields4}
C_\perp(\sigma)&=&{}^{gr}C_\perp(\sigma)+{}^{mat}C_\perp(\sigma) 
\nn\\
&=& \frac{1}{\kappa}\left[g^{-\frac{1}{2}}(g_{ac}g_{bd}-\frac{1}{d-1}g_{ab}g_{cd})p^{ab}p^{cd}(\sigma) + g^{\frac{1}{2}} R(\sigma)\right] +{}^{mat}C_\perp(\sigma) \q . \q\q
\ea
Here $g=\text{det}(g_{ab})$ is the determinant and $R$ the Ricci scalar for the spatial metric $g_{ab}$. For a non--derivative coupling the matter part ${}^{mat}C_\perp$ may depend on the metric $g_{ab}$ but not on the gravitational momentum $p^{ab}$.

There may arise additional constraints for the matter fields, but we will assume that these are already solved and one is working on the reduced phase space with respect to these matter constraints.

General Relativity is a totally constrained system, that is the Hamiltonian of the theory is a linear combination of the constraints
\ba\label{fields5}
H[N,N^a]=\int_\Sigma \big(N(\sigma) C_\perp(\sigma)+N^a(\sigma)C_a(\sigma) \big)\td^d \sigma
\ea 
where % $\td^d \sigma$ is the coordinate volume element of $\Sigma$ and
 $N$ and $N^a$ are called lapse and shift function. Apart from the condition that the lapse has to be strictly positive, the lapse and shift function can be chosen arbitrarily. The constraint algebra is given by 
\ba \label{diracalgebra}
\{H[M,M^a],H[N,N^a]\}= \,H[M^bN_{,b}-N^bM_{,b} \, \, ,\,\,\,M^bN^a_{,b}-N^bM^a_{,b} +g^{ab}(MN_{,b}- NM_{,b} )]  \q .\nn\\
\ea
Because of the appearence of the inverse metric $g^{ab}$ this algebra is a constraint algebra with structure functions.

Since the system is totally constrained, we know that the Hamiltonian equations of motion 
\ba\label{fields6}
\frac{\partial}{\partial s}g_{ab}(s,\sigma) &=& \{g_{ab}(s,\sigma),H[N,N^a]\} \nn \\
\frac{\partial}{\partial s}p^{ab}(s,\sigma) &=& \{p^{ab}(s,\sigma),H[N,N^a]\} 
\ea
and similarly for the matter fields describe the evolution with respect to an a priori unphysical time parameter $s$.

Assume that one has found a complete solution to the equations of motion (\ref{fields6}) with some choice of lapse and shift function. We now want to reconstruct the space--time metric $\gamma_{\mu\nu}$ on the space--time manifold, in the following denoted by $\cs$. So far we have just the $s$--dependent spatial metric $g_{ab}(s,\sigma)$ on the $d$-dimensional manifold $\Sigma$, the lapse and shift functions $N(s,\sigma),N^a(s,\sigma)$ and the conjugated momentum $p^{ab}(s,\sigma)$. 

In order to reconstruct the space--time metric, one has  to embed the surface $\Sigma$ into the manifold $\cs$ for each value of the parameter $s$. To this end introduce a coordinate system $(z^\mu)_{\mu=0,\ldots,d}$ on $\cs$. The embeddings can then be written as  
\ba\label{fields7}
Z_s: \Sigma &\rightarrow& \cs \nn \\
       \sigma &\mapsto&  Z_s(\sigma)\equiv (Z^\mu(s,\sigma))_{\mu=0}^d
\ea
where $(Z^\mu(s,\sigma))_{\mu=0}^d $ are the coordinates for the point $Z_s(\sigma)$. The image of $Z_s$ is the embedded hypersurface $\Sigma_s \subset \cs$.

The embeddings (\ref{fields7}) induce $d$ independent vector fields $Z_a^\mu, a=1,\ldots,d$ on $\cs$, which are tangential to the hypersurfaces $\Sigma_s$:
\ba \label{fields8}
Z^\mu_a(z):=Z^\mu_{,a}(s,\sigma)_{| z^\nu = Z^\nu(s,\sigma)} \q .
\ea 
%Here a comma followed by an index denotes the partial derivative with respect to the corresponding coordinate, in this case $\sigma^a$.  

The derivative of the embeddings (\ref{fields7}) with respect to the parameter $s$ defines the space--time deformation vector
\ba\label{fields9}
N^\mu(z):=\frac{\partial Z^\mu}{\partial s}(s,\sigma)_{|z^\nu = Z^\nu(s,\sigma)} \q .
\ea
The embeddings (\ref{fields7}) should be such, that the $(d+1)$ vector fields $(N^\mu,Z^\mu_a),a=1,\ldots,d$ form a basis of the tangent space to $\cs$. One can then define the space--time metric $\gamma_{\mu\nu}$ by giving its matrix elements with respect to this basis:
\begin{xalignat}{2}\label{fields10}
&\gamma_{\mu\nu}N^\mu N^\nu(z)=(-N^2+N^cN_c)(z) &   &\gamma_{\mu\nu}N^\mu Z_b^\nu(z)=N_b(z)  \nn \\
&\gamma_{\mu\nu}Z_a^\mu N^\nu(z)\,=N_a(z) & &\gamma_{\mu\nu}Z_a^\mu Z_b^\nu (z)\;=g_{ab}(z)
\end{xalignat}
where $N(z),N^a(z)$ are the lapse and shift functions in the Hamiltonian (\ref{fields5}), 
%at $z^\mu=Z^\mu(s,t)$, 
$g_{ab}(z)$ is the spatial metric and $N_a=N^bg_{ab}$. Here $z$ refers to the point with coordinates $z^\mu=Z^\mu(s,\sigma)$.

Hence given the family of spatial metrics $g_{ab}(s,\sigma)$ satisfying (\ref{fields6}) and the lapse and shift functions $N,N^a(s,\sigma)$ one can reconstruct the space--time metric $\gamma_{\mu\nu}$.

 In order to get the $z^\mu$-coordinate expression for the space--time metric we introduce the future-pointing unit normal vector to the hypersurface $\Sigma_s$ 
\ba\label{fields11}
n^\mu(z)=\frac{1}{N}(N^\mu-N^a Z^\mu_a)(z) \q .
\ea
%where the lapse and shift function $(N,N^a)(z)$ at $z^\mu=Z^\mu(s,t)$ are the ones from the Hamiltonian (\ref{fields7}). 
The inverse space--time metric $\gamma^{\mu\nu}$ can then be written as
\ba\label{fields12}
\gamma^{\mu\nu}(z)=Z^\mu_a Z^\nu_a g^{ab}(z)-n^\mu n^\nu(z)  \q 
\ea
where $g^{ab}(z)$ is the inverse of $g_{ab}$ at $z^\mu=Z^\mu(s,\sigma)$. From the inverse space--time metric one can calculate the space--time metric $\gamma_{\mu\nu}(z)$. 

The last equation in (\ref{fields10}) means that the space--time metric $\gamma_{\mu\nu}$ induces a spatial geometry on $\Sigma_s$ which coincides with the geometry defined by $g_{ab}(s,\sigma)$. 
 Also, the lapse and shift functions have a geometrical meaning. Because of (\ref{fields11}) we have that the deformation vector $N^\mu$, which connects two consecutive hypersurfaces $\Sigma_s$ and $\Sigma_{s+\varepsilon}$ is given by
\ba\label{fields13}
N^\mu=Nn^\mu+N^aZ^\mu_a \q .
\ea
That is the lapse function $N$ gives the proper distance $\varepsilon N$ between the two consecutive hypersurfaces $\Sigma_s$ and $\Sigma_{s+\varepsilon}$ measured in normal direction. The shift vector $\varepsilon N^a$ gives the tangential deformation that is applied to the points of $\Sigma$ if the embedding is changed from $Z_s(\sigma)$ to $Z_{s+\varepsilon}(\sigma)$.  

Here we began with the $s$--dependent spatial metric fields $g_{ab}(s,\sigma)$ on $\Sigma$ and the lapse and shift functions $N(s,\sigma)$ and $N^a(s,\sigma)$. We reconstructed the space--time metric field $\gamma_{\mu\nu}$ using the family of embeddings (\ref{fields7}). If we had used another family of embeddings $Z'_s$ we would have ended up with another space--time metric field $\gamma'_{\mu'\nu'}$. But this metric field would by definition induce the same geometry on the surfaces $Z'_s(\Sigma)$ as the metric field $\gamma_{\mu\nu}$ on the surfaces $Z_s(\Sigma)$ and the foliation $Z'_s(\Sigma)$ would be still characterized by the same lapse and shift functions. Indeed one can understand the change from $\gamma_{\mu\nu}$ to $\gamma'_{\mu'\nu'}$ as the effect of a coordinate transformation 
\ba\label{fields14}
z^\mu \mapsto z^{\mu'}={Z'}_s(s,\sigma)_{|Z^\mu(s,\sigma)=z^\mu}
\ea
which also changes the coordinates of the surfaces $\Sigma_s$ from $z^\mu=Z^\mu(s,\sigma)$ to $z'^{\mu'}=Z'^{\mu'}(s,\sigma)$. 

However one can go the other way around and consider a fixed space--time metric field $\gamma_{\mu\nu}$ on $\cs$ satisfying the Einstein equations. A family of embeddings $Z_s$ as in (\ref{fields7}) can then be used to define $s$--dependent spatial  metrics on $\Sigma$ by
\ba\label{fields15}
g_{ab}(s,\sigma)=Z^\mu_a Z^\nu_b \gamma_{\mu\nu}(z)_{|z^\mu=Z^\mu(s,\sigma)} \q .
\ea 
Also, the equations (\ref{fields9}) and (\ref{fields10}) determine lapse and shift functions $N(s,\sigma)$ and $N^a(s,\sigma)$. Moreover the conjugated momentum $p^{ab}$ is defined by 
\ba\label{fields16}
p^{ab}(s,\sigma)=g^{\frac{1}{2}}(K^{ab}-g^{ab}K^c_c)(s,\sigma)
\ea
where we raise and lower indices with $g^{ab}$ and $g_{ab}$ respectively. $K_{ab}$ is the extrinsic curvature of $\Sigma_s$:
\ba\label{fields16a}
K_{ab}(s,\sigma)=Z^\mu_a Z^\nu_b \nabla_\mu n_\nu (z)_{|z^\mu=Z^\mu(s,\sigma)} \q .
\ea
Here $\nabla$ is the covariant differential on $\cs$ associated to the space--time metric $\gamma_{\mu\nu}$.

In general a space--time tensor field (for instance a matter field) on $\Sigma_s\subset \cs$ can be pulled back to a spatial tensor field on $\Sigma$ in the following way
\ba\label{inf3}
{t^{a\cdots}}_{b\cdots}(s,\sigma)=Z^a_\mu\cdots Z_b^\nu \cdots {t^{\mu\cdots}}_{\nu\cdots}(z)_{|z^\rho=Z^\rho(s,\sigma)}
\ea
where 
\ba\label{inf4}
Z^a_\mu=g^{ab}\gamma_{\mu\nu}Z^\nu_b  \q .
\ea
%and as in (\ref{fields10}) the spatial metric $g_{ab}$ is the pull-back of the space--time metric $\gamma_{\mu\nu}$
%\ba\label{inf5}
%g_{ab}=Z^\mu_a Z^\nu_b \gamma_{\mu\nu} \q .
%\ea
%$g^{ab}$ is the inverse to $g_{bd}$. 

 But in pulling back tensor fields to $\Sigma$ we will always project out the normal components of these tensor-fields, because we have $Z^\mu_a n_\mu=Z_\mu^a n^\mu=0$ by definition. One can regain these components by first projecting tensor components onto the normal vector $n^\mu$ and then pulling back the resulting expression:
\ba\label{inf6}
{t^{a\cdots\perp\cdots}}_{b\cdots\perp\cdots}(s,\sigma)=(-1)^\cn\,
Z^a_\mu\cdots n_\rho \cdots Z_b^\nu\cdots n^\kappa \cdots {t^{\mu\cdots\rho\ldots}}_{\nu\cdots\kappa\cdots}(z)_{|z^\rho=Z^\rho(s,\sigma)}
\ea   
where $\cn$ is the number of contractions of lower or upper indices with the normal vector or the normal covector. We can then regain for instance a space--time vector as $t^\mu=Z^\mu_a t^a+n^\mu t^\perp$.
The two operations of pulling back and projecting onto the normal vector give the fields we are working with in the canonical formalism.
%\footnote{D-invariance}.

Now if we change the embeddings $Z_s$ to $Z'_s$ but leave the space--time metric field $\gamma_{\mu\nu}$ fixed, the induced spatial metrics and conjugated momenta will also change as well as the lapse and shift functions. 
%Since the field $\gamma_{\mu\nu}$ satisfies the Einstein equations, these new canonical data have to fulfill the equations of motion (\ref{fields6}) with respect to the Hamiltonian $H[N',{N'}^a]$.  

In particular consider a family of embeddings of $\Sigma$ with:
\ba\label{fields17}
{Z}^\mu(\varepsilon,\sigma)=Z^\mu(0,\sigma)+ \varepsilon \Lambda(\sigma) n^\mu(z)_{|z^\nu=Z^\nu(0,\sigma)}+ \varepsilon \Lambda^a(\sigma)Z^\mu_a(0,\sigma)+O(\varepsilon^2)  \q .
\ea

One can show \cite{hyperspace, 7.3} that the induced canonical data on $\Sigma$ for parameter value $s=\varepsilon$ are given by
\ba\label{fields18}
g_{ab}(\varepsilon,\sigma)&\simeq& g_{ab}(0,\sigma)+\{g_{ab}(0,\sigma), H[\Lambda,\Lambda^a]\}_{|s=0}+O(\varepsilon^2) \nn \\
p^{ab}(\varepsilon,\sigma)&\simeq&p^{ab}(0,\sigma)+ \{p^{ab}(\sigma), H[\Lambda,\Lambda^a]\}_{|s=0}+O(\varepsilon^2)   \q 
\ea
with
\ba\label{fields19}
H[\Lambda,\Lambda^a]=\int_\Sigma \big(\Lambda C_\perp(\sigma)+\Lambda^aC_a(\sigma)\big) \td^d \sigma \q .
\ea
Hence one can say that the Hamiltonian constraint $C_\perp$ generates deformations of the embedding of $\Sigma$ in normal direction to the embedded surface and that the diffeomorphism constraints $C_a$ generate deformations in the tangential directions, i.e. spatial diffeomorphisms. We want to remark that equations (\ref{fields18}) are only valid if the canonical data $g_{ab}(s,\sigma)$ and $p^{ab}(s,\sigma)$ are induced from a space--time metric satisfying the Einstein equations.

\section{Complete Observables Associated to Space--Time Scalars}\label{completescalars}

%From a lagrangian viewpoint the gauge transformations of general relativity are given by the diffeomorphism group on the space--time manifold $\cs$ and its induced action on the space--time fields. As was noted in the last section in the canonical formalism the diffeomorphism constraints generates spatial diffeomorphisms and the Hamiltonian constraints generates deformations of the spatial hypersurface in normal directions. 

It is often suggested \cite{BergmannHandbuch, deWitt, RovPartObs} to construct the following space--time diffeomorphism invariant quantity: Choose $(d+1)$ space--time scalar fields $\psi^K,K=0,\ldots,d$ and calculate the value of a $(d+2)$--th scalar field $\phi$ at that point in space--time at which the fields $\psi^K$ assume the values $\xi^K$. 

Hence, in order to define these observables, one needs to specify only $(d+1)$ parameter values $\xi^K$ and not infinitely many parameters $\tau^K(\sigma),\sigma\in \Sigma$, as is the case for the complete observables in the canonical formalism. 

The question arises whether one can define the same kind of observables also in the canonical formalism. In the following we will show, that this is indeed the case. To this end we will use the ideas sketched in section \ref{completefields} and therefore look for partial observables, which are already invariant with respect to almost all constraints. Then it may indeed happen, that the corresponding complete observable does only depend on $(d+1)$ parameters and not on the parameter fields $\tau^K(\sigma),\sigma\in \Sigma $. Therefore let us contemplate the question, how one can decide in the canonical formalism, whether a certain phase space quantity can be reconstructed as a space--time scalar or not.

I.e. given a spatial scalar $\psi$ on $\Sigma$, built from the canonical fields. The equations of motions for the latter will result in a field $\psi(s,\sigma)$ on $\Rl\times\Sigma$. This can be mapped to a field $\psi(z)$ on the space--time manifold $\cs$ by the family of embeddings $Z_s$ (see (\ref{fields7})). Can one decide whether $\psi(z)$ is also a space--time scalar or rather something else, like the projection of a space--time vector onto the normal vector fields to the surfaces $\Sigma_s$?     

Indeed there is a necessary and sufficient condition for $\psi$ to be reconstructable as a space--time scalar. Likewise there are conditions for all other kinds of space--time tensors, summarized in the reconstruction theorem of Kucha\v{r}, \cite{hyperspace}. For the moment we will only consider the condition for space--time scalars, since this will give as the kind of partially invariant partial observables we are looking for: A field $\psi$ built from the canonical fields and behaving as a scalar under spatial diffeomorphisms of $\Sigma$, is reconstructable as a space--time scalar if and only if we have for every point $\sigma\in \Sigma$
\ba\label{inf8}
%\{\psi(\sigma),\int_\Sigma N^a C_a(\sigma')\td^d\sigma'\}\simeq 0 \q \text{and} \q
\{\psi(\sigma),\int_\Sigma N\, C_\perp \,(\sigma')\,\td^d \sigma'\}\simeq 0 \q
\ea
for all lapse functions $N(\sigma')$ with $N(\sigma)=0$. That is the Poisson brackets of a space--time scalar field with the Hamiltonian constraints (and also with the diffeomorphism constraints, see below) have to be ultra--local.

One can prove this assertion by realizing that infinitesimal gauge transformations with lapse functions that vanish at the point $\sigma$ deform the position of the hypersurface $\Sigma_s$ in such a way, that the point $\sigma$ is mapped to the same space--time point as before. That is, the hypersurface is tilted but the point $Z_s(\sigma)$ is held fixed. A space--time scalar should only depend on the point $Z_s(\sigma)$ but not on the shape of the hypersurface $\Sigma_s$ containing this point. In contrast to that, the projection of a space--time vector to the normal unit vector should depend on the shape of the hypersurface, since the normal vector itself depends on the shape of the hypersurface, see appendix \ref{appendix1}.  

A field $\psi$ behaves as a spatial scalar if it exhibits the following Poisson brackets with the diffeomorphism constraints:
 \ba\label{inf9}
\{\psi(\sigma),\int_\Sigma N^a C_a(\sigma')\td^d\sigma'\}=N^a(\sigma) \psi_{,a}(\sigma) \q .
\ea
So the Poisson brackets with diffeomorphism constraints smeared with shift functions $N^a$ which vanish at the point $\sigma$ also vanish.

In the following we will call all canonical fields satisfying equations (\ref{inf8}) and (\ref{inf9}) space--time scalars. As we have seen, space--time scalars exhibit exactly those properties, which we need in order to apply the reasonig from section \ref{completefields}.

That is choose as clock variables $(d+1)$ fields $T^K,\,K=0,\ldots,d$, satisfying the equations (\ref{inf8}, \ref{inf9}). Then we can write for the Poisson brackets between the latter and the constraints
\ba\label{inf10}
{A^K}_\perp(\sigma,\sigma')=\{T^K(\sigma),C_\perp (\sigma')\} &\simeq&\{T^K(\sigma),C_\perp [1]\}\, \delta(\sigma, \sigma')=:{B^K}_\perp(\sigma)\delta(\sigma, \sigma') \nn \\
{A^K}_a(\sigma,\sigma')=  \{T^K(\sigma),C_a (\sigma')\} &=& T^K_{,a}(\sigma) \, \delta(\sigma, \sigma')\q\q\q\q=:{B^K}_a(\sigma)\delta(\sigma, \sigma')
\ea
where we defined
\ba\label{inf11}
C_\perp [1]:=\int_\Sigma C_\perp(\sigma') \td^d \sigma'   \q .
\ea

In the following we will use lower case letters from the middle of the alphabet to denote the set of indices $\{\perp,a=1,\ldots,d\}$. The weakly Abelian constraints at a point $\sigma$ are just linear combinations of the old constraints at the same point:
\ba\label{inf16}
\tilde C_K(\sigma)={(B^{-1})^j}_K (\sigma)\, C_j (\sigma)  \q .
\ea

Now, choose as partial observable $f$ a spatial scalar field $\psi$ evaluated at a point $\sigma^*$, that is $f=\psi(\sigma^*)$, where $\psi$ satisfies the requirements (\ref{inf8}, \ref{inf9}). Then we have also for the new constraints
\ba\label{inf17}
\{\psi(\sigma^*),\tilde C_K (\sigma) \}\simeq \{\psi(\sigma^*),\tilde C_K [1] \} \, \delta(\sigma^*,\sigma) \q 
\ea
with $\tilde C_K[1]=\int_\Sigma \tilde C_K(\sigma')\,\td \sigma'$.\footnote{One can also use $\tilde C_K[\Lambda_{(K)}]:=\int_\Sigma \Lambda_{(K)}(\sigma) \tilde C_K(\sigma) \td^d\sigma$ with $\Lambda_{(K)}(\sigma^*)=1$.}

% Let us apply the functional differential equation (\ref{}) for the associated complete observable $F_{[\psi(\sigma^*);T]}$. If we change the parameter values $\tau^K(\sigma)$ by
%\ba\label{fields18}
%\Delta \left(\tau^K(\sigma)\right)=\varepsilon \Lambda^K(\sigma)
%\ea
%the complete observable changes to
%\ba\label{fields19}
%F_{[\psi(\sigma^*);T]} (\tau+\Delta\tau,x)\simeq F_{[\psi(\sigma^*);T]}(\tau,x) + \varepsilon\Lambda^K (\sigma^*)\, \, F_{[\{\psi(\sigma^*),\tilde C_K (1) \};T]}(\tau,x) +O(\varepsilon^2) \q  .
%\ea
%Hence, the complete observable does not depend on the parameter values $\tau^K(\sigma)$ for all points $\sigma\in \Sigma$ with $\sigma \neq \sigma^*$. 

Since the new constraints $\tilde C_K$ are weakly Abelian, the iterated Poisson brackets of $\psi(\sigma^*)$ with the new constraints will also involve only delta functions and we obtain for the complete observable associated to $\psi(\sigma^*)$:
\ba\label{fields21a}
F_{[\psi(\sigma^*);T]}(\tau,\cdot)
&=&
\sum_{r=0}^\infty \frac{1}{r!}\{\cdots\{\psi(\sigma^*),\tilde C_{K_1}[1]\},\cdots, \tilde C_{K_r}[1]\} 
\nn\\
 &&\q\q\q\q
(\tau^{K_1}(\sigma^*)-T^{K_1}(\sigma^*))\cdots(\tau^{K_r}(\sigma^*)-T^{K_r}(\sigma^*)) \q . \q\q
\ea

As for the covariant construction involving $(d+2)$ space--time scalar fields mentioned at the beginning of this section the right hand side of equation (\ref{fields21a}) does depend only on the $(d+1)$ parameters $\tau^K:=\tau^K(\sigma^*)$ and not on the infintely many other values $\tau^K(\sigma),\sigma\neq\sigma^*$.

%Furthermore, the value of the complete observable does not depend on the choice of the point $\sigma^*\in \Sigma$, that is, we have
%\ba\label{fields19a}
%F_{[\psi(\sigma^*);\, T]}(\tau,\cdot)\simeq F_{[\psi(\sigma^{**});\,T]}(\tau',\cdot)
%\ea
%if the parameter fields $\tau^K$ and ${\tau'}^K$ satisfy $\tau^K(\sigma^*)={\tau'}^K(\sigma^{**})$ for all $K=0,\ldots,d$.

Remember that the constraints generate deformations in the embedding of the hypersurface $\Sigma$ into the space--time manifold $\cs$.
Hence $F_{[\psi(\sigma^*);T]}(\tau,\cdot)$ gives the value of $\psi(\sigma^*)$ on that family of embedded hypersurfaces for which $T^K(\sigma^*)=\tau^K$. I.e. these embeddings have to map the point $\sigma^*$ to the space--time point $z^*$ characterized by $T^K(z^*)=\tau^K$. In other words $F_{[\psi(\sigma^*);T]}(\tau,\cdot)$ gives the value of the space--time scalar $\psi$ at that point in the space--time manifold where the scalar fields $T^K$ are equal to $\tau^K$. In this way the fields $T^K$ serve as a {\it physical} coordinate system.  

Note that in the power series (\ref{fields21a}) we have only to deal with the $(d+1)$ constraints $\tilde C_K[1]$ and not with the infintely many constraints $\tilde C_K(\sigma),\sigma\in \Sigma$. In section \ref{solvediffeo} we will reduce the number of constraints we have to deal with to one.

\section{The Metric as a Space--Time Scalar}\label{spacetimemetric}

We have seen that if one uses space--time scalars as partial observables, that is phase space functionals $\psi(\sigma)$ satisfying the equations (\ref{inf8}, \ref{inf9}), the associated complete observable will only depend on $(d+1)$ parameters and not on infinitely many parameters. 

Now the question arises whether there are enough phase space functionals with these properties. Of course, if gravity is coupled to matter fields, one can construct space--time scalars out of these matter fields. One can also use the (space--time) curvature tensor and construct space--time scalars from it by contraction. For a four--dimensional space--time one can use the so called Bergmann--Komar invariants \cite{BergmannHandbuch}, which are four independent space--time scalars built from the space--time curvature tensor. These invariants can be expressed through the canonical data $(g_{ab},p^{ab})$ and matter fields, if one uses the equation of motion \cite{BergmannHandbuch}.

Assume that one has chosen $(d+1)$ space--time scalars $T^K,\,K=0,\ldots,d$, for instance the Bergmann--Komar invariants or scalars built from matter fields. Then one can even express the inverse space--time metric $\gamma^{\mu\nu}$ in the $T^{K}$ frame by
\ba\label{fields22} 
\gamma^{KL}(z)=T^K_{,\mu}T^L_{,\nu}\,\gamma^{\mu\nu}(z)  \q .
\ea
Now, if one applies a space--time diffeomorphism to $\gamma^{KL}$ the behaviour of $\gamma^{\mu\nu}$ as a contravariant tensor is balanced by the covariant vectors $T^K_{,\mu}$. Hence each component of $\gamma^{KL}$ behaves as a space--time scalar. Using equation (\ref{fields12}) these can also be expressed through canonical data and independently of lapse and shift functions in the following way
 \ba\label{fields23}
\gamma^{KL}(z)&=&\left[ T^K_{,\mu}T^L_{,\nu}(z)(Z^\mu_a Z^\nu_b g^{ab}-n^\mu n^\nu)(s,\sigma)\right]_{|z^\mu=Z^\mu(s,\sigma)} \nn \\
&=&  \left[ T^K_{,a}T^L_{,b}g^{ab}(s,\sigma)-n^\mu T^K_{,\mu} \, n^\nu T^L_{,\nu}(s,\sigma)\right]_{|z^\mu=Z^\mu(s,\sigma)} \nn \\
&=& \left[ T^K_{,a}T^L_{,b}g^{ab}(s,\sigma)-\{T^K,C_\perp[1]\}\,\{T^L,C_\perp[1]\}(s,\sigma)\right]_{|z^\mu=Z^\mu(s,\sigma)} \q .
\ea
Here we used, that
\ba\label{fields24a}
T^K_{,a}(s,\sigma)=Z^\mu_a(s,\sigma) T^K_{,\mu}(z)_{|z^\mu=Z^\mu(s,\sigma)}
\ea
in the second line of (\ref{fields23}) and the equations of motion 
\ba\label{fields25b}
n^\mu T^K_{,\mu}(z)=
% \cl_{n^\mu}T^K\,(z)=
\{T^K,C_{\perp}[1]\}(s,\sigma)_{|z^\mu=Z^\mu(s,\sigma)}
\ea
in the third line. (Remember that $C_\perp$ generated deformation of the hypersurface in normal direction.) From now on we will abbreviate $\{T^K,C_{\perp}[1]\}$ with $-T^K_{,\perp}$, where the minus sign appears because of the convention in (\ref{inf6}).

The space--time metric $\gamma_{KL}$ inverse to $\gamma^{LM}$, is given by
\ba\label{fields25a}
\gamma_{KL}=
{(B^{-1})^a}_K{(B^{-1})^b}_L \,g_{ab}- 
{(B^{-1})^\perp}_K{(B^{-1})^\perp}_L  \q  
\ea
where ${(B^{-1})}^j_K$ is the inverse to the matrix ${B^K}_j$ introduced in (\ref{inf10}).

In the following we will show that the phase space functional
\ba\label{fields24}
\gamma^{KL}(\sigma):=T^K_{,a}T^L_{,b}g^{ab}(\sigma)-T^K_{,\perp}T^L_{,\perp}(\sigma)
\ea
satisfies indeed the requirements in order to be reconstructable 
as a space--time scalar. That is, we will show that the Poisson 
brackets of this quantity with the constraints $
(C_\perp(\sigma'),C_a(\sigma'))$ and therefore also with the 
weakly Abelian constraints $\tilde C_K(\sigma')$ does involve only 
delta functions $\delta(\sigma,\sigma')$ at least on the constraint
 hypersurface. Since the new constraints are weakly Abelian
 this property holds at least weakly also for the 
Poisson bracket of $\gamma^{KL}$ with the smeared constraints 
$\tilde C_K[1]$. This gives us further phase space quantities, for
 which formula (\ref{fields21a}) is applicable. 

For the calculation of the Poisson brackets of $\gamma_{KL}$ with the constraints, we will begin with the diffeomorphism constraints $C_c(\sigma')$. Since $T^K$ are spatial scalars the first summand in (\ref{fields24}) has to be also a spatial scalar, i.e. we have
\ba\label{fields25}
\{ T^K_{,a}T^L_{,b}g^{ab}(\sigma), C_c (\sigma')\}=\left[T^K_{,a}T^L_{,b}g^{ab}\right]_{,c}(\sigma)\,\delta(\sigma,\sigma') \q .
\ea
This also holds  for the normal derivatives $-T^K_{,\perp}(\sigma)$ in (\ref{fields24}). More explicitly, we use the Jacobi identity and the fact that the smeared Hamiltonian constraint $C_\perp(1)$ commutes with the diffeomorphism constraints $C_c(\sigma')$ (see \ref{diracalgebra}):
\ba\label{fields26}
\{-T^K_{,\perp}(\sigma),C_c(\sigma')\} &=& \{\{T^K,C_\perp[1]\},C_c(\sigma')\} \nn \\
&=&  \{\{T^K,C_c(\sigma')\},C_\perp [1]\} \nn \\
&=& \{T^K_{,c}(\sigma),C_\perp [1]\}\,\delta(\sigma,\sigma') \nn \\
&=&\delta(\sigma,\sigma')\,\, \frac{\partial}{\partial \sigma^c} \{T^K(\sigma),C_\perp [1]\} \nn \\
&=:&- T^K_{,\perp c} (\sigma) \delta(\sigma,\sigma') \q .
\ea
Here we applied in the fourth line the identity
\ba\label{fields27}
\{ \frac{\partial}{\partial \sigma^a} \phi(\sigma), \psi(\sigma')\}=\frac{\partial}{\partial \sigma^a} \{\phi(\sigma), \psi(\sigma')\} \q .
\ea
%which holds for any fields $\phi,\psi$. 

Hence we have
\ba\label{fields28}
\{\gamma^{KL}(\sigma),C_c(\sigma')\}={\gamma^{KL}}_{,c}(\sigma)\,\delta(\sigma,\sigma')
\ea
and therefore $\gamma^{KL}(\sigma)$ behaves indeed as a spatial scalar.

For the Poisson bracket of $\gamma^{KL}$ with the Hamiltonian constraints $C_\perp(\sigma')$ we need the Poisson brackets between $T^K_{,a}$ and the Hamiltonian constraints
\ba\label{fields29}
\{T^K_{,a}(\sigma),C_\perp(\sigma')\}=-\frac{\partial}{\partial \sigma^a}\left[T^K_{,\perp}(\sigma)\, \delta(\sigma,\sigma')\right]
\ea
and between $g^{ab}$ and the Hamiltonian constraints 
\ba\label{fields30}
\{g^{ab}(\sigma),C_\perp(\sigma')\}=-2 g^{-\frac{1}{2}} (p^{ab}-\frac{1}{d-1} g^{ab}p) (\sigma) \delta(\sigma,\sigma')=\{{g^{ab}}(\sigma),C_\perp[1]\}\,\delta(\sigma,\sigma') \q .\q\q
\ea

Furthermore, in order to calculate the Poisson brackets of $-T^K_{,\perp}$ with the Hamiltonian constraints we will use again the Jacobi identity and the constraint algebra (\ref{diracalgebra}), namely 
\ba\label{fields31zusatz}
\{C_\perp[1],C_\perp(\sigma')\}=-\frac{\partial}{\partial {\sigma'}^a} (g^{ab} C_b)(\sigma') \q ,
\ea
so that we can write
\ba\label{fields31}
\{-T^K_{,\perp}(\sigma), C_\perp(\sigma')\}
&=&  
\{\{T_K,C_\perp[1]\},C_\perp(\sigma')\} \nn \\
&=&
\{\{T_K(\sigma),C_\perp(\sigma')\},C_\perp[1]\}-\{\{C_\perp[1],C_\perp(\sigma')\},T^K(\sigma)\} \nn \\
&=&
\{-T^K_{,\perp}(\sigma),C_\perp[1]\}\,\delta(\sigma,\sigma')+\{\frac{\partial}{\partial {\sigma'}^a} (g^{ab} C_b)(\sigma'),T^K(\sigma)\} \nn \\
&=&
\{-T^K_{,\perp}(\sigma),C_\perp[1]\}\,\delta(\sigma,\sigma')+
\frac{\partial}{\partial {\sigma'}^a}\left[ C_b(\sigma')\,\{g^{ab}(\sigma'),T^K(\sigma)\}\right]+ \nn \\
&&\q\q\q\q\q\q\q\q\q\q\q\;
\frac{\partial}{\partial {\sigma'}^a}\left[g^{ab}(\sigma')\,\{C_b(\sigma'),T^K(\sigma)\}\right] \nn \\
&\simeq& \{-T^K_{,\perp}(\sigma),C_\perp[1]\}\,\delta(\sigma,\sigma')-\frac{\partial}{\partial {\sigma'}^a}\left[g^{ab}(\sigma') T^K_{,b}(\sigma)\, \delta(\sigma,\sigma')\right] \nn \\
&\simeq&\{-T^K_{,\perp}(\sigma),C_\perp[1]\}\,\delta(\sigma,\sigma')-g^{ab}(\sigma) T^K_{,b}(\sigma) \frac{\partial}{\partial {\sigma'}^a}\,\delta(\sigma,\sigma')
\ea
where in the last line we used the delta function identity
\ba\label{fields32}
\frac{\partial}{\partial {\sigma'}^b} \left[ \psi(\sigma')\phi(\sigma)\,\delta(\sigma,\sigma')\right]
&=&\frac{\partial}{\partial {\sigma'}^b} \left[ \psi(\sigma)\phi(\sigma)\,\delta(\sigma,\sigma')\right] \nn \\
&=& \psi(\sigma)\phi(\sigma)\frac{\partial}{\partial {\sigma'}^b}\delta(\sigma,\sigma') \q .
\ea
This equation can be proved by integrating both sides with test functions $\Lambda(\sigma)$ and $\Lambda'(\sigma')$.  

Finally in order to put the results of (\ref{fields29}, \ref{fields30}, \ref{fields31}) together, we utilize another delta function identity which can be proved in a similar manner as (\ref{fields32}):
\ba\label{fields33}
\frac{\partial}{\partial {\sigma'}^b}\delta(\sigma,\sigma')=-\frac{\partial}{\partial {\sigma}^b}\delta(\sigma,\sigma') \q .
\ea
The derivatives of the delta functions in (\ref{fields29}) and (\ref{fields31}) then cancel each other and we will end with
\ba\label{fields34}
\{\gamma^{KL}(\sigma),C_\perp (\sigma')\} &\simeq&\bigg[ -T^K_{,\perp a}T^L_{,b}\,g^{ab}-T^K_{,a}T^L_{,\perp b}\, g^{ab} +T^K_{,a}T^L_{,b} \, \{g^{ab},C_\perp[1]\} \nn \\
&& \q - T^L_{,\perp}\,\{T^K_{,\perp},C_\perp[1]\}-T^K_{,\perp}\,\{T^L_{,\perp},C_\perp[1]\} \bigg](\sigma)\,\delta(\sigma,\sigma') \nn \\
&\simeq& \{\gamma^{KL}(\sigma),C_\perp[1]\} \,\delta(\sigma,\sigma') \q .
\ea
This finishes the proof that $\gamma^{KL}$ and therefore also $\gamma_{KL}$ can be reconstructed as a space--time scalar.

These considerations allow the construction of all kinds of space--time diffeomorphism invariant quantities connected to metric properties as observables in the canonical formalism. For instance the space--time volume of a certain space--time region specified by the values of the fields $T^K$ can be expressed as the integral over the corresponding $\tau^K$--parameters over the complete observable associated to the determinant of $\gamma_{KL}$. Quantities which involve derivatives of the metric, as for instance curvatures, can be constructed by using the derivatives with respect to the parameters $\tau^K$ of the complete observables associated to the matrix elements of $\gamma^{KL}$. These coincide with the complete observables associated to the phase space function $\{\gamma^{KL}(\sigma), \tilde C_K[1]\}$, which is also a space--time scalar.

In appendix \ref{appendix1} we show how one can express general space--time tensor fields in the $T^K$--coordinate system and that the so gained $T^K$--components behave as space--time scalars, i.e. have ultra--local Poisson brackets with the constraints.

\section{Solving the Diffeomorphism Constraints}\label{solvediffeo}

Using space--time scalars as partial observables , 
 we managed to reduce
 the number of constraints we have effectively to deal with
 to the $(d+1)$ constraints $\tilde C_K[1]$. That is, only 
these constraints appear in the power series (\ref{fields21a}).

A further reduction can be obtained if one introduces
complete observables with respect to the diffeomorphism constraints
 $C_a(\sigma),\sigma\in\Sigma,a=1,\ldots,d$. This was also done in \cite{dust} for the example of gravity coupled to an incoherent dust (albeit with another terminology). Let us explain this idea in more detail:

 In \cite{bd1} it was shown that one can construct complete observables `in stages', i.e. one can first look for phase space functions which are invariant under a subalgebra of the constraint algebra (e.g. the diffeomorphism constraints) and then use these partially invariant phase space functions as partial observables in order to construct complete observables which are invariant under all the constraints. 
 
 If one applies this strategy to the diffeomorphism and the Hamiltonian constraints, this means that one has first to construct spatially diffeomorphism invariant phase space functions. Then one uses these phase space functions as clock variables and partial observables in order to construct complete observables with respect to the Hamiltonian constraints. One does need as many clock variables as there are constraints left after the first step, i.e. as there are Hamiltonian constraints, that is just one clock variable for each point $\sigma\in \Sigma$. The proof in \cite{bd1} shows, that if one constructs a complete observable associated to the diffeomorphism invariant clock variables and a diffeomorphism invariant partial observable with respect to the Hamiltonian constraints, this complete observable is not only invariant under these Hamiltonian constraints but also under the diffeomorphism constraints, which were not involved in the construction of the complete observable. This statement holds despite the fact, that e.g. the Poisson bracket of two Hamiltonian constraints is equal to a diffeomorphism constraint, i.e. that the Hamiltonian constraints do not close among themselves.

 One method to complete the first step, i.e. to find spatial diffeomorphism invariant partial observables, is to construct complete observables with respect to the diffeomorphism group. One could then use these partially complete observables as partial observables in order to construct fully complete observables with respect to the Hamiltonian constraints. However one has then to work with two very different types of phase space functions: on the one hand with the partially complete observables which will for instance not be labelled by the spatial points $\sigma\in \Sigma$ but by the parameter fields $\tau^A(\sigma),A=1,\ldots,d$ (in special cases by just $d$ parameters $\tau^a$) and on the other hand we have the Hamiltonian constraints $C_\perp(\sigma)$ which are labelled by the spatial points $\sigma\in \Sigma$.
 
We will therefore replace the Hamiltonian constraints $C_\perp(\sigma),\, \sigma\in \Sigma$ by another set of constraints which are also gained as partially complete observables with respect to the diffeomorphism group. Hence these new constraints are Hamiltonian constraints invariant under spatial diffeomorphism and therefore if one computes complete observables with respect to these constraints one works entirely with functions that are invariant under spatial diffeomorphisms.

But first we have to show, that the so introduced new Hamiltonian constraints are equivalent to the old ones. To this end consider a first class constraint algebra $\Fc$ with a subalgebra $\Fc_1$. The constraints in this subalgebra will be denoted by $C_a$ where $a$ is here a possibly continuous index in an index set $\ca$. The remaining constraint set will be denoted by $\Fc_2$ and the constraints in this set by $C_j$ with an index $j$ from an index set $\cj$. Choose a set of clock variables $T_a,\,a\in \ca$ such that the equations $T_a=\tau_a\; \forall a\in \ca$ provide good gauge conditions for some fixed real numbers $\tau_a,\,a\in\ca$.\footnote{i.e. the chosen gauge must be accessible from an arbitrary point on the constraint hypersurface and the gauge conditions must fix the gauge completely, that is there is no gauge transformation other than the identity, that preserves the gauge} This condition will ensure that the partially complete observables are well--defined. 

Define the partially complete observables $D_{[C_j:\,T_c]}(\tau_c,\cdot)$ with respect to the constraint set $\Fc_1$ and associated to the clock variables $T_c,\,c\in \ca$ and the constraints $C_j,\,j\in\cj$ (and with the fixed parameters $\tau_a,\,a\in\ca$). Here we use the symbol $D$ instead of $F$ in order to indicate that we compute the complete observable with respect to the subalgebra $\Fc_1$. We have to show that the vanishing of the constraints $D_{[C_j:\,T_c]}(\tau_c,\cdot)$ at points of the submanifold $\cc_1$, defined by the vanishing of the constraints in $\Fc_1$, is equivalent to the vanishing of the original constraints $C_j,\,j\in\cj$ at such points. 

 On the one hand if all the constraints $C_j,\,j\in\cj$ vanish at a point $x\in\cc_1$ then also the constraints $D_{[C_j:\,T_c]}(\tau_c,\cdot),\,j\in\cj$ vanish at this point. The reason for that is, that $D_{[C_j:\,T_c]}(\tau_c,\cdot)$ is at least linear in the constraints $C_a\in \Fc_1$ or $C_j\in \Fc_2$ as can be seen by using the power series (\ref{july9}) for complete observables and the fact that the constraint algebra $\Fc$ is first class, i.e. Poisson brackets of constraints are at least linear in the constraints.

 On the other hand if all the constraints $D_{[C_j;\,T_c]}(\tau_c,\cdot)$ vanish at a point $x\in \cc_1$ they will vanish on the complete $\Fc_1$--orbit through the point $x$. In particular they will vanish at that point $y$ in this orbit where $T_a(y)=\tau_a$ for all $a\in\ca$. But at this point we have $C_j(y)=D_{[C_j;T_c]}(\tau_c,y)=0$. Hence also the constraints $C_j$ vanish at the point $y$, i.e. this point and also the $\Fc_1$--orbit through this point and in particular the point $x$ are positioned inside the constraint hypersurface $\cc$, where the constraints $C_j,\,j\in\cj$ and the constraints $C_a,\,a\in\ca$ vanish. 

Therefore the vanishing of the constraints $D_{[C_j;T_c]}(\tau_c,\cdot),\,j\in\cj$ is equivalent to the vanishing of the constraints $C_j,\,j\in\cj$: Both constraint sets are equivalent if completed with the constraint algebra $\Fc_1$. But now $\Fc_1$ is an ideal in the algbra $\Fc_1 \cup \{D_{[C_j;T_c]}(\tau_c,\cdot),\,j\in\cj\}$.

Given a $\Fc_1$--invariant function $f$ and $\Fc_1$--invariant clock variables $T_j,\,j\in\cj$ one can therefore compute the complete observable either with respect to the constraints $C_j$ or with respect to the constraints $D_{[C_j;T_c]}(\tau_c,\cdot)$. The result of both procedures is (weakly) the same, since both complete observables are invariant under the constraints and coincide on the ($\Fc_2$--)gauge fixing $\{T_k=\tau_k,\,k\in \cj\}$.

%The calculation with the $\Fc_1$--invariant constraint set $\{D_{[C_j;T_c]}(\tau_c,\cdot)\}_{j=1}^m$ might be easier if one already knows the partielly complete observables associated to all phase space coordinates. According to (\ref{poissondirac}) one can then express the Poisson bracket between two partially complete observables as a partially complete observable:
%\ba
%\{D_{[f;T_c]}(\tau_c,\cdot),D_{[g;T_c]}(\tau_c,\cdot)\}=D_{[\{f,g\}^D;T_c]}(\tau_c,\cdot) \q .
%\ea
%Here $\{\cdot,\cdot\}^D$ is the Dirac bracket with respect to the constraint set $\Fc_1$ and the gauge fixings $\{T_a=\tau_a\}_{a=1}^n$. We will apply both methods for calculating complete observables (with the constraints $C_j$ and the constraints $D_{[C_j,T_c]}(\tau_c,\cdot)$) in sections \ref{completescalars} and \ref{solvediffeo}.

%%%%%%%%%%%%%%%%%%%%%%%%%%%%%%%%%%%%%%%%%%%%%%%%%%%%%%%
%%%%%%%%%%%%%%%%%%%%%%%%%%%%%%%%%%%%%%%%%%%%%%%%%%%%%%
%%%%%%%%%%%%%%%%%%%%%%%%%%%%%%%%%%%%%%%%%%%%%%%%%%%%%%%

 Let us next consider the construction of partially complete observables with respect to the diffeomorphism group. Afterwards we will come back to the partially complete observables associated to the Hamiltonian constraints.

 To begin with we choose $d$ clock variable fields $T^A(\sigma),A=1,\ldots,d$ and another field $\psi(\sigma)$. We have to assume that the fields $T^A$ define a good\footnote{I.e. the map $\ct:=(T^A):\Sigma \rightarrow (T^A)(\Sigma) \subset\Rl^d$ should be bijective for each phase space point on the constraint hypersurface $\cc_D$ defined by the vanishing of all diffeomorphism constraints. Furthermore we will assume that the image $(T^A)(\Sigma) \subset\Rl^d$ is the same for each phase space point in $\cc_D$.} coordinate system on $\Sigma$, this means in particular that $\text{det}(T^A_{,a})$ should not vanish. 
%This assumptions will in general restrict the phase space region in which the following considerations are valid.
We will demand that the fields $T^A$ behave as space time scalars according to the requirements (\ref{inf8}, \ref{inf9}).

We can then define the partially complete observable $D_{[\psi(\sigma^*);T^A]}(\cy)$ with respect to the diffeomorphism constraints and associated to the phase space functional $\psi(\sigma^*)$. Here we renamed the parameter field $\tau^A(\sigma)$ into $\cy=(\cy^A(\sigma))_{A=1}^d$ and omitted the dependence on the phase space point $x$. 

If the field $\psi(\sigma)$ behaves as a spatial scalar, i.e. if we have
\ba\label{diff1}
\{\psi(\sigma), C_b(\sigma')\}=\psi_{,b}(\sigma)\delta(\sigma,\sigma') \q ,
\ea
then the complete observable $D^\cy_{\psi(\sigma^*)}$ will only depend on the $d$ parameters $Y^A:=\cy^A(\sigma^*)$ and not on the values the parameter fields $\cy^A$ assume at the other points of $\Sigma$. This can be understood by the same reasoning as used in section \ref{completescalars} to argue that complete observables associated to space--time scalars depend only on $d+1$ parameters.

%That is we can interpret the complete observable $D^\cy_{\psi(\sigma^*)}$ as the value of the field $\psi$ at that point $\sigma \in\Sigma$ at which the three fields $T^A$ give the values $T^A(\sigma)=\cy^A(\sigma^*):=Y^A$. If $\psi$ is a spatial scalar this defines a phase space dependent coordinate transformation for the field $\psi$:
%\ba\label{diff2}
%D^\cy_{\psi(\sigma^*)}=\psi(S(Y))
%\ea
%where $S$ is the inverse mapping to $\cy$, i.e. $S(\cy(\sigma))=\sigma$. Formula (\ref{diff2}) shows that $D^\cy_{\psi^*}$ does indeed only depend on the $d$ parameters $Y^A=\cy^A(\sigma^*)$. 

Now any spatial (densitized) tensor can be made into a spatial scalar by multiplying it with appropriate factors of the Jacobian matrix $(\partial T^A/\partial \sigma^a)_{a,A=1}^d$, its inverse and its determinant. For instance for the spatial metric and the conjugated momentum we can define the quantities
\ba\label{diff3}
g_{AB}(Y) &:=& D_{[g_{ab} {S^a}_A {S^b}_B(\sigma^*);T^A]}(\cy) \\
p^{AB}(Y)&:=&  D_{[p^{ab} T^A_{,a} T^B_{,b} \,\text{det}({S^c}_C)\,(\sigma^*);T^A]}(\cy)
\ea
where $({S^a}_A(\sigma))_{a,A=1}^d$ is the inverse matrix to the matrix $(T^A_{,b}(\sigma))_{b,A=1}^d$ and the fields $\cy^A$ satisfy $\cy^A(\sigma^*)=Y^A$. 

In the following we will assume that the fields $\{T^A\}_{A=1}^d$ are part of the canonical coordinates and we will denote the conjugated momenta by $\{\Pi_A\}_{A=1}^d$. Furthermore we will suppose that there is a space--time scalar field $T^0$ such that $\{T^0(\sigma),C_\perp(1)\}$ does not vanish. This field will be used as clock variable with respect to the Hamiltonian constraints. We define the partially complete observables associated to these variables in the following way
\ba\label{diff4}
T^0(Y)&:=& D_{[T^0(\sigma^*);T^A]}(\cy) \\
%\Pi_0(Y)&:=& D^\cy_{\Pi_0 \,\text{det}({S^c}_C)\,(\sigma^*)} \\
\Pi_A(Y)&:=& D_{[\Pi_A \, \text{det}({S^c}_C)\,(\sigma^*);T^A]}(\cy) \q 
\ea
where again $Y^A=\cy^A(\sigma^*)$.

In order to define partially complete observables associated to the Hamiltonian constraints we first multiply these by $\text{det}({S^c}_C)$ since the Hamiltonian constraints are densities of weight one. We then have spatially diffeomorphism invariant Hamiltonian constraints
\ba\label{bjuly1}
C_\perp(Y):=D_{[\text{det}({S^c}_C)C_\perp(\sigma^*);\,\,T^A]}(\cy)
\ea 
which will also depend only on the $d$ parameters $Y^A=\cy^A(\sigma^*)$ and not on the whole parameter fields $\cy^A(\sigma),\sigma\in\Sigma$. 

We have shown above that the constraints $\{D_{[\text{det}({S^c}_C)C_\perp(\sigma);T^A]}(\cy),\,\sigma\in\Sigma\}$ are equivalent to the original set of constraints $\{C_\perp(\sigma)\,\sigma\in\Sigma\}$ if one sublements both sets with the diffeomorphism constraints. (The determinant factor does not matter, since it was assumed that it does not vanish.) Now we have that $D_{[\text{det}({S^c}_C)C_\perp(\sigma);T^A]}(\cy)=C_\perp(Y)$ where $Y=\cy(\sigma)$. Since we have assumed that the fields $T^A$ provide a good coordinate system we have that 
\ba\label{bjuly2} 
\{D_{[\text{det}({S^c}_C)C_\perp(\sigma);T^A]}(\cy),\,\,\sigma\in\Sigma\}=\{C_\perp(Y),\,\,(Y^A)_{A=1}^d\in(T^A)_{A=1}^d(\Sigma)\subset \Rl^d\} \q . 
\ea

Hence we can use the constraint set $\{C_\perp(Y),\,\,(Y^A)_{A=1}^d\in(T^A)_{A=1}^d(\Sigma)\subset \Rl^d\}$ and the other partially complete observables in order to compute fully complete observables. 
 For this we need to specify the Poisson brackets between two partially complete observables. In \cite{bd1} it was shown that the Poisson bracket between two complete observables associated to the phase space functions $f$ and $g$ is given by the complete observable associated to the Dirac bracket $\{f,g\}^D$ of the two partial observables. Here the Dirac bracket is with respect to the gauge fixing where all the clock variables are equal to some constants. That is we have  
\ba\label{diff5}
\{D_{[\phi(\sigma^*);\,T^A]}(\cy),D_{[\psi(\sigma^{**});\,T^A]}(\cy)\}=D_{[\{\phi(\sigma^*),\psi(\sigma^{**})\}^D\,;\, T^A]} (\cy)
\ea
where $\{\cdot,\cdot\}^D$ is the Dirac bracket with respect to the diffeomorphism constraints $\{C_b(\sigma)\}_{b=1}^d,\sigma\in \Sigma$ and the gauge fixings $\{T^A(\sigma)=\cy^A(\sigma);\sigma\in \Sigma\}$. This equation holds on the submanifold $\cc_D$ of phase space, defined by the vanishing of all the diffeomorphism constraints. Therefore in the following all equations need just to hold on this hypersurface $\cc_D$.

The Dirac bracket is given by
\ba\label{diracbrjuly}
\{\phi(\sigma^*),\psi(\sigma^{**})\}^D &=&\{\phi(\sigma^*),\psi(\sigma^{**})\} 
- \int_\Sigma \{\phi(\sigma^*), {S^b}_A C_b(\sigma)\}\,\{T^A(\sigma), \psi(\sigma^{**})\}\,\td^d\sigma  
\nn\\
&&\q\q\q\q\q\q\q + \int_\Sigma \{\phi(\sigma^*),T^A(\sigma)\}\,\{{S^b}_A C_b(\sigma), \psi(\sigma^{**})\}\,\td^d\sigma  \q\q
\ea
and coincides with the usual Poisson bracket if the two phase space functions involved do not depend on the momenta $\{\Pi_A\}_{A=1}^d$, that is if they have vanishing Poisson brackets with the clock variables $\{T^A\}_{A=1}^d$. However we can replace every phase space function by a phase space function not depending on these momenta if one solves the diffeomorphism constraints $C_b$: since the clock variables are scalars, these have to be of the form
\ba\label{diff6}
C_b=\frac{1}{\alpha_A}\Pi_A T^A_{,b}+ C^{(R)}_b
\ea
where $C^{(R)}_b$ does not depend on the momenta $\{\Pi_A\}_{A=1}^d$. In particular the diffeomorphism constraints are linear in these momenta and hence can be easily solved for these:
\ba\label{diff7}
\Pi_A =- \alpha_A{S^b}_A C^{(R)}_b \q .
\ea
Since $D_{[C_b(\sigma^*)\,; T^A]}(\cy)=0$ on $\cc_D$, the partially complete observable associated to $\Pi_A$ will coincide with the observable associated to $-\alpha_A{S^b}_A C^{(R)}_b$ on $\cc_D$. Hence before calculating Poisson brackets between partially complete observables, we replace the momenta $\Pi_A$ by the expressions $-\alpha_A{S^b}_A C^{(R)}_b$. If this is done the Dirac brackets are just given by the usual Poisson brackets.

Let us consider the Poisson brackets between $g_{AB}$ and $p^{CD}$ as an example:
\ba\label{diff8}
\{\,D_{[g_{ab}{S^a}_A {S^b}_B(\sigma^*);T^A]}(\cy),D_{[p^{cd} T^C_{,c} T^D_{,d}\text{det}({S^c}_C) (\sigma^{**});T^A]}(\cy)\,\}\!\!
&=&\!\! \kappa D_{[\text{det}({S^c}_C) (\sigma^{**})\delta(\sigma^*,\sigma^{**})\delta_{(A}^C\delta_{B)}^D;T^A]}(\cy)\nn \\
&=&\!\!\kappa\delta_{(A}^C\delta_{B)}^D[\text{det}(\cy^C,c)(\sigma^{**})]^{-1}\,\, \delta(\sigma^*,\sigma^{**}) \nn\\
&=&\!\!\kappa\delta_{(A}^C\delta_{B)}^D \delta(Y',Y'')  \q 
\ea
where we used that $D_{[T^B(\sigma);T^A]}(\cy)=\cy^B(\sigma)$ and defined ${Y'}^A:=\cy^A(\sigma^*)$ and ${Y''}^A:=\cy^A(\sigma^{**})$.
%The term $\text{det}({S^c}_C) (\sigma^{**})\delta(\sigma^*,\sigma^{**})$ is a spatial scalar in both arguments $\sigma^*,\sigma^{**}$ (with the convention that the delta function is a spatial scalar in its first argument and a density in its second argument). 
Hence the fields $g_{AB}$ and $p^{AB}$ are canonically conjugated to each other. This holds also for all the other partially complete observables associated to canonically conjugated fields (except for the clock variables $T^A$ and their conjugated momenta).

In summary the Poisson brackets between partially complete observables are structurally the same as the Poisson brackets between the original fields with the exceptions of terms that depend on the momenta $\{\Pi_A\}_{A=1}^d$. But these can be replaced by using (\ref{diff7}). In particular a field  $\psi(\sigma)$ being a space--time scalar will retain this property, that is at least weakly we will have
\ba\label{diff9}
\{\psi(Y),C_\perp(Y')\}\sim \delta(Y,Y') \q .
\ea 
This follows if we apply formula (\ref{diracbrjuly}) in order to calculate the Dirac bracket between the field $\psi(\sigma^*)$ and the constraint $\text{det}({S^c}_C)C_\perp(\sigma^{**})$:
\ba
\{\psi(\sigma^*), \text{det}({S^c}_C) C_\perp(\sigma^{**})\}^D 
&\simeq &
\text{det}({S^c}_C)(\sigma^{**}) 
\bigg(
\{\psi(\sigma^*), C_\perp(\sigma^{**})\} - 
\nn\\
&& \q\q
\int_\Sigma  \{\psi(\sigma^*), {S^b}_A C_b(\sigma)\} 
               \{T^A(\sigma),C_\perp(\sigma^{**})\}\, \td^d \sigma + 
\nn\\
&& \q\q \int_\Sigma \{\psi(\sigma^*),T^A(\sigma)\}\{{S^b}_A C_b(\sigma), C_\perp(\sigma^{**})\} \td^d \sigma  
\bigg)\q .
 \q \nn \\ 
\ea
The first two summands on the right hand side do not contain derivatives of delta functions because all Poisson brackets appearing there are at least weakly proportional to delta functions. Furthermore because of the first class property of the constraints the last summand vanishes weakly. Hence we can write
\ba
\{\psi(\sigma^*), \text{det}({S^c}_C) C_\perp(\sigma^{**})\}^D 
&\simeq &
\text{det}({S^c}_C)(\sigma^{**}) \delta(\sigma^*,\sigma^{**})
\bigg(
\{\psi(\sigma^*), C_\perp[1]\} -
\psi_{,b} {S^b}_A  T^A_{,\perp}(\sigma^*)\bigg)\; .
\nn\\
\ea
Now, as one can check by direct calculation (see for instance (\ref{fields26})), the two summands in the big brackets behave as spatial scalars. That is, the partially complete observable associated to these summands will only depend on the $d$ parameters $Y^A=\cy^A(\sigma^*)$. A calculation similar to (\ref{diff8}) shows that
\ba
\{\psi(Y),C_\perp(Y')\} \simeq  D_{(\{\psi(\sigma^*), C_\perp[1]\} -
\psi_{,b} {S^b}_A  T^A_{,\perp}(\sigma^*))\,;\, T^A]}(Y^A)
\, \delta(Y,Y')
\ea
which proves the assertion (\ref{diff9}).

Therefore if one computes complete observables associated to such a field $\psi(Y)$  and a clock variable field $T^0(Y)$ arising also from a space--time scalar field $T^0(\sigma)$ one can apply the same reasoning as in section \ref{completescalars}: Define the weakly Abelian Hamiltonian constraints
\ba\label{diff10}
\tilde C(Y):= B^{-1} (Y) C_\perp(Y)
\ea
where $B(Y)$ is determined by $\{T^0(Y),C_\perp(Y')\}=B(Y)\delta(Y,Y')$. The iterated Poisson brackets of $\psi(Y)$ with the constraints $\tilde C(Y'),\tilde C(Y''),\ldots$ will contain only delta functions and no derivatives of delta functions. Therefore we can write for the power series of the complete observable associated to the phase space functional $\psi(Y^*)$
\ba\label{diff11}
F_{[\psi(Y^*);T^0]}(\tau,\cdot)=\sum_{r=0}^\infty \frac{1}{r!}\{\psi(Y^*),\tilde C[1]\}_r\,\, (\tau^0(Y^*)-T^0(Y^*))^r
\ea
where $\tilde C[1] :=\int_{\cy(\Sigma)} \tilde C(Y) \td^d Y$.

Hence for the computation of the complete observable associated to a field $\psi(Y)$ satisfying (\ref{diff9}) we are reduced to the consideration of just one constraint $\tilde C[1]$.

%%%%%%%%%%%%%%%%%%%%%%%%%%%%%%%Abelian
%%%%%%%%%%%%%%%%%%%%%%%%%%%%%%%

\section{Abelian Diffeomorphism Invariant Hamiltonian Constraints}\label{AbelianConstraints}

Under the assumption that there exists such clock variable fields $T^A(\sigma),\,\sigma\in\Sigma$ as explained in the last section we managed to replace the original Hamiltonian constraints by diffeomorphism invariant Hamiltonian constraints. 

The Poisson bracket of two smeared Hamiltonian constraints $C_\perp[N]$ and $C_\perp[N']$ is equal to a smeared diffeomorphism constraint, see equation (\ref{diracalgebra}). Therefore one might speculate that the diffeomorphism invariant Hamiltonian constraints Poisson commute with each other. Let us check this by a direct calculation. 
 The Poisson bracket of two diffeomorphism invariant Hamiltonian constraints is given by
 \ba\label{cjuly1}
 \{C_\perp(Y),C_\perp(Y')\}=D_{[\{\text{det}({S^c}_C) C_\perp (\sigma^*), \text{det}({S^c}_C) C_\perp (\sigma^{**})\}^D\, ;\,\,T^A]}(\cy)
 \ea
 where $Y=\cy(\sigma^*)$ and $Y'=\cy(\sigma^{**})$ and from here on all equations hold modulo diffeomorphism constraints.\footnote{However, note the following point. Since the spatial diffeomorphism constraints define an algebra with structure constants in contrast to an algebra with structure functions, it is possible to define the constraints $C_\perp(Y)$ as phase space functions which are invariant under the diffeomorphism constraints on the whole phase space and not just on the hypersurface $\cc_D$ defined by the vanishing of the diffeomorphism constraints. If this is done the Poisson bracket of two such constraints $C_\perp(Y)$ and $C_\perp(Y')$ has to be also diffeomorphism invariant, i.e. if it involves the diffeomorphism constraints then only in a diffeomorphism invariant combination.}

 According to (\ref{diracbrjuly}) the Dirac bracket appearing on the right hand side of the last equation can be calculated to
\ba\label{cjuly2}
\{\text{det}({S^c}_C) C_\perp (\sigma^*), \text{det}({S^c}_C) C_\perp (\sigma^{**})\}^D\!\!\!\!
&=&
\{C_\perp (\sigma^*),C_\perp (\sigma^{**})\}\text{det}({S^c}_C)(\sigma^*) \text{det}({S^c}_C)(\sigma^{**}) +
\nn\\
&&
\{C_\perp (\sigma^*),\text{det}({S^c}_C)(\sigma^{**}) \}\text{det}({S^c}_C)(\sigma^*)C_\perp (\sigma^{**})+
\nn\\
&&\{\text{det}({S^c}_C)(\sigma^*),C_\perp (\sigma^{**})\}C_\perp (\sigma^*)\text{det}({S^c}_C)(\sigma^{**})+
\nn\\
&&\left( \text{det}({S^c}_C)C_\perp\right)_{,b}{S^b}_A(\sigma^*) \text{det}({S^c}_C)T^A_{,\perp}(\sigma^{**})\,\delta(\sigma^*,\sigma^{**})-
\nn\\
 &&\text{det}({S^c}_C)T^A_{,\perp}(\sigma^{*})\left( \text{det}({S^c}_C)C_\perp\right)_{,b}{S^b}_A(\sigma^{**})\,\delta(\sigma^*,\sigma^{**})
 \nn\\
 &=& 
 \{C_\perp (\sigma^*),\text{det}({S^c}_C)(\sigma^{**}) \}\text{det}({S^c}_C)(\sigma^*)C_\perp (\sigma^{**})+
\nn\\
&&\{\text{det}({S^c}_C)(\sigma^*),C_\perp (\sigma^{**})\}C_\perp (\sigma^*)\text{det}({S^c}_C)(\sigma^{**}) \q 
\nn\\
\ea
where the first term on the right hand side does not appear in the next equation because it is proportional to a diffeomorphism constraint. The remaining Poisson brackets are given by
\ba\label{cjuly3}
\{C_\perp (\sigma^*),\text{det}({S^c}_C)(\sigma^{**}) \}
&=&
-\text{det}({S^c}_C) {S^a}_A(\sigma^{**})\{C_\perp(\sigma^*),T^A_{,a}(\sigma^{**})\}
\nn\\
&=&
-\text{det}({S^c}_C) {S^a}_A(\sigma^{**})\frac{\partial}{\partial {\sigma^{**}}^a}\left( T^A_{,\perp}(\sigma^{**}) \delta(\sigma^{**},\sigma^*)\right)
\ea
where in the last line we used equation (\ref{fields29}). Hence for the Dirac bracket (\ref{cjuly2}) we are left with
\ba\label{cjuly4a}
\{\text{det}({S^c}_C) C_\perp (\sigma^*), \text{det}({S^c}_C) C_\perp (\sigma^{**})\}^D\!\!\!\!
&=&
-\text{det}({S^c}_C) C_\perp T^A_{,\perp}{S^a}_A(\sigma^{**})
\text{det}({S^c}_C)(\sigma^*)\,\times\nn\\
&& \q\q\q
\frac{\partial}{\partial {\sigma^{**}}^a}\,\delta(\sigma^{**},\sigma^*)\nn\\
&&+\text{det}({S^c}_C) C_\perp T^A_{,\perp}{S^a}_A(\sigma^{*})
\text{det}({S^c}_C)(\sigma^{**})\,\times\nn\\
&& \q\q\q
\frac{\partial}{\partial {\sigma^{*}}^a}\,\delta(\sigma^{*},\sigma^{**}) \q .
\ea
That gives for the Poisson brackets (\ref{cjuly1})
\ba\label{cjuly4}
\{C_\perp(Y),C_\perp(Y')\}&=&-T^A_{,\perp} C_\perp(Y') \frac{\partial}{\partial {Y'}^A} \delta(Y',Y)
+T^A_{,\perp} C_\perp(Y) \frac{\partial}{\partial {Y}^A} \delta(Y,Y') \q .
\ea
%smear with test functions in $Y,Y'$
The constraints $C_\perp(Y)$ are non--Abelian because of the Poisson bracket between the determinant $\text{det}({S^c}_C)(\sigma^{**})$ and the constraint $C_\perp(\sigma^*)$ which contains a derivative of a delta function. However one might try to multiply the constraint  $C_\perp(\sigma^*)$ with another non--vanishing density of weight $(-1)$ in order to get a spatial scalar and such that this density has ultra--local Poisson brackets with the constraints. Such a density would be $g^{-\frac{1}{2}}$. Let us therefore define
\ba\label{cjuly5}
C^{abel}(Y)=D_{[g^{-\frac{1}{2}}C_\perp(\sigma^*);\,T^A]}(\cy)=g^{-\frac{1}{2}}(Y)C_\perp(Y)
\ea
where $Y=\cy(\sigma^*)$. (Remember that $g^{-\frac{1}{2}}(Y)=D_{[\text{det}(T^C_{,c})g^{-\frac{1}{2}}(\sigma^*);\,T^A]}(\cy)$.) Indeed we have for the Poisson brackets of these constraints
\ba\label{cjuly6}
\{g^{-\frac{1}{2}}C_\perp(\sigma^*),g^{-\frac{1}{2}}C_\perp(\sigma^{**})\}^D\!
=\!\!\!\!\!\!&&
\{g^{-\frac{1}{2}}(\sigma^*),C_\perp(\sigma^{**})\}C_\perp(\sigma^*)g^{-\frac{1}{2}}(\sigma^{**})+
\nn\\
&&
\{C_\perp(\sigma^*),g^{-\frac{1}{2}}(\sigma^{**})\}g^{-\frac{1}{2}}(\sigma^*)C_\perp(\sigma^{**})
-\nn\\
&&
\int_\Sigma \left(g^{-\frac{1}{2}}C_\perp\right)_{,a}(\sigma^*) {S^a}_A(\sigma)\delta(\sigma^*,\sigma) 
\{ T^A(\sigma), g^{-\frac{1}{2}}C_\perp(\sigma^{**}) \}\,\td^d\sigma -
\nn\\
&&
\int_\Sigma  \{g^{-\frac{1}{2}}C_\perp(\sigma^{*}), T^A(\sigma)\}\,
\left(g^{-\frac{1}{2}}C_\perp\right)_{,a}(\sigma^{**}) {S^a}_A(\sigma)\delta(\sigma^{**},\sigma)\, \td^d\sigma \, .\nn\\
\ea
The first two terms on the right hand side of the last equation cancel each other because the Poisson brackets involved are proportional to delta functions. The last two terms cancel each other if the Poisson bracket 
\ba\label{cjuly7}
\{ T^A(\sigma), g^{-\frac{1}{2}}C_\perp(\sigma^{**}) \}=-T^A_{,\perp}(\sigma)g^{-\frac{1}{2}}(\sigma^*)\delta(\sigma,\sigma*)
+\{T^A(\sigma),g^{-\frac{1}{2}}(\sigma^{**}) \}C_\perp(\sigma^{**})
\ea
is ultra--local, i.e. if the last term on the right hand side of (\ref{cjuly7}) vanishes or at least does not contain derivatives of delta functions. This is the case, if the $T^A$ are matter fields. 

That is, if $\{T^A(\sigma),g^{-\frac{1}{2}}(\sigma^{**}) \}\sim\delta(\sigma,\sigma^*)$ or if this expression vanishes the constraints $C^{abel}(Y)$ are Abelian (modulo diffeomorphism invariant terms that vanish on $\cc_D$). We can also choose to work with these constraints in order to calculate fully complete observables. Note however, that if the clock variables $T^0(Y)$ Poisson commute with $g^{-\frac{1}{2}}(Y')$ the constraints $\tilde{C}^{abel}(Y):=(\{T^0(Y),C^{abel}[1]\})^{-1} C^{abel}(Y)$ coincide with the constraints $\tilde{C}(Y)$ defined in (\ref{diff10}).

We want to remark that diffeomorphism invariant Abelian Hamiltonian constraints were defined before, for instance in \cite{dust} for gravity coupled to incoherent dust. However there Abelianess is reached by another method: first one solves the Hamiltonian and the diffeomorphism constraints for the momenta $\Pi_K$ conjugated to the matter fields $T^K,K=0,1,\ldots,d$. This results in constraints of the form $C_K=\Pi_k+h_K$ where the phase space functions $h_K$ do not depend on the momenta $\Pi_K$. Such constraints are Abelian but they will typically involve square roots, since the original Hamiltonian constraints are quadratic in (some of) the momenta. One can then define partially complete observables with respect to the diffeomorphism constraints and associated to the new constraints $C_0(\sigma),\sigma\in\Sigma$ (multiplied by the determinant of the inverse of $(T^A,_{,a}(\sigma))_{a,A=1}^d)$. These constraints will still be Abelian, because the constraints $C_0(\sigma)$ are independent from the momenta $\Pi_A,A=1,\ldots,d$. Our constraints have the advantage that it is not necessary to solve the constraints for some of the momenta and hence they do not invovle square roots but will have a similar structure as the original constraints, see the example in section \ref{example}.

\section{Coupling gravity to scalar fields}\label{example}

In this section we will consider gravity (in four space--time dimensions) coupled to four scalar fields $\{T^K\}_{K=0}^3$. These four scalar fields will serve as clock variables. 

The diffeomorphism and Hamiltonian constraints for this system are given by
\ba\label{diff12}
C_b &=&{}^{gr}C_b+{}^{mat}C_b   \nn\\
C_\perp &=& {}^{gr}C_\perp  +{}^{mat}C_\perp 
\ea
with the matter contributions
\ba \label{diff14}
{}^{mat}C_b 
&=& 
\sum_K {}^{(K)}C_b
=
\sum_K \frac{1}{\alpha_K}\Pi_K T^K_{,b} 
\nn\\
{}^{mat}C_\perp
&=&   
\sum_K {}^{(K)}C_\perp  
= 
\sum_K \frac{1}{2\alpha_K}\left( g^{-\frac{1}{2}} (\Pi_K)^2 + g^{\frac{1}{2}} g^{ab} T^K_{,a} T^K_{,b} +2g^{\frac{1}{2}}V_{(K)}(T^L)\right)\, \, . \q\q  
\ea
Here $\Pi_K$ are the momenta conjugated to the scalar fields $T^K$ such that $\{T^K(\sigma),\Pi_L(\sigma')\}=\alpha_K \delta^K_L\,\delta(\sigma,\sigma')$ where $\alpha_K$ are coupling constants for the matter fields. $V_{(K)}(T^L)$ are potentials for the matter fields, depending on the scalar fields $T^L$.

Let us calculate the weakly Abelian constraints. First, we will do that for arbitrary couplings and an arbitrary choice of space--time scalar fields as clock variables $T^K$. The matrix ${B^K}_j$ defined in (\ref{inf10}) can be written as
\ba\label{exjuly1}
{B^J}_j= \left(
   \begin{array}{ll}
 -T^0_{,\perp} & T^0_{,a}  \\
 -T^A_{,\perp} & T^{A}_{,a}   \end{array}\right)
=\left(
   \begin{array}{ll}
   -T^0_{,\perp} & T^0_{,b}\, {S^b}_B \\
   -T^A_{,\perp}   & \delta^{A}_B  \end{array}\right)
   \left( \begin{array}{ll}
     1 & 0  \\
     0 & T^B_{,a}   \end{array}\right)
\ea  
where ${S^b}_B$ is the inverse to $T^B_{,a}$. This allows us to invert the matrix (\ref{exjuly1}) in the following way:
\ba\label{exjuly2}
{(B^{-1})^j}_J  &=& \frac{1}{E} 
\left(  \begin{array}{ll}
 1 & 0 \\
 0 & {S^a}_B  \end{array}\right)
\left( \begin{array}{ll}
 1\q & -T^0_{,b}\, {S^b}_A \\
 T^B_{,\perp}\q & E\,\delta^B_A -T^B_{,\perp}\,  T^0_{,b}\, {S^b}_A 
\end{array}\right) \\
&=& 
\frac{1}{E}    \left(  \begin{array}{ll}
1 \q &  -T^0_{,b}\,{S^b}_A \\
{S^a}_B\,T^B_{,\perp}\q   & 
E\,{S^a}_A-{S^a}_B\, T^B_{,\perp}\,  T^0_{,b}\, {S^b}_A
\end{array}\right)
\ea
where
\ba\label{exjuly3}
E= -T^0_{,\perp} +T^0_{,a} {S^a}_A T^A_{,\perp}  \q .
\ea
The weakly Abelian constraints are then given by $\tilde C_K = {(B^{-1})^j}_K C_j$, e.g. for $K=0$ we have
\ba\label{exjuly4}
\tilde C_0 (\sigma)= \frac{1}{E} (C_\perp + T^A_{,\perp} {S^b}_A C_b)(\sigma) \q .
\ea
For the coupling of gravity to four scalar fields (\ref{exjuly4}) becomes
\ba\label{exjuly5}
\tilde C_0(\sigma) = \left( g^{-\frac{1}{2}}\Pi_0 - g^{-\frac{1}{2}} T^0_{,a} {S^a}_A \Pi_A\right)^{-1}\left(C_\perp-g^{-\frac{1}{2}}\Pi_A {S^b}_A C_b\right)(\sigma) \q .
\ea

For the matrix elements of the inverse space--time metric $\gamma^{KL}(\sigma):=T^K_{,a}T^L_{,b}g^{ab}(\sigma)-T^K_{,\perp}T^L_{,\perp}(\sigma)$ we will get
\ba\label{diff19}
\gamma^{KL}(\sigma)=T^K_{,a}T^L_{,b} g^{ab}-g^{-1}  \Pi_K \Pi_L (\sigma) \q ,
\ea
i.e. the space--time metric in the $T^K$ frame involves the conjugated momenta to the fields $T^K$.

%%%%%%%%%%%%%%%%%%%%%%%%%%%%%%%%%%%%%%%%%%%%
%%%%%%%%%%%%%%%%%%%%%%%%%%%%%%%%%%%%%%%%%%%%
%%%%%%%%%%%%%%%%%%%%%%%%%%%%%%%%%%%%%%%%%%%%%
Let us turn to the methods developed in section \ref{solvediffeo}. The clock variables do not satisfy all the assumptions made in this section, nevertheless it might be helpful to use the diffeomorphism invariant Hamiltonian constraints defined there, in order to calculate Dirac observables, which are at least well defined in some parts of phase space, that is in parts in which the clock variables give good coordinates in the sense of section \ref{solvediffeo}.

The spatial diffeomorphism invariant Hamiltonian constraints are given by
\ba\label{diff15}
C_\perp(Y) 
&=&
\frac{1}{\kappa}g^{-\frac{1}{2}}(p_{AB}p^{AB}-\frac{1}{2}p^2)(Y)+
\frac{1}{\kappa}g^\frac{1}{2}R(Y)+
\frac{1}{2\alpha_0}g^{-\frac{1}{2}} (\Pi_0)^2(Y)+
\nn\\
&&
\frac{1}{2\alpha_0} g^\frac{1}{2}g^{AB}T^0_{,A}T^0_{,B}(Y)+
%\nn\\
%&&
\frac{1}{2} g^{-\frac{1}{2}}(Y)\sum_{A=1}^3 \frac{1}{\alpha_A}(\Pi_A)^2(Y)+
\nn\\
&&
\frac{1}{2} g^{\frac{1}{2}}(Y)\sum_{A=1}^3 \frac{1}{\alpha_A}g^{AA}(Y) +
%\nn \\
%&& \q\q\q\q\q\q\q\q
g^\frac{1}{2} \sum_{K=0}^3 \frac{1}{\alpha_K}V_{(K)}(T^0(Y),Y^B)\, .\q\q\q
\ea
Here we used the notation $g(Y)=\text{det}(g_{AB}(Y))$, $T^0_{,A}(Y)=(T^0_{,b}S^b_A)(Y)$ and so on. The momenta $\{\Pi_A(Y)\}_{A=1}^3$ are abbreviations for 
\ba\label{diff16}
\Pi_A(Y) &=&-\alpha_AD^\cy_{\text{det}({S^c}_C) S^{b}_A({}^{gr}C_b+{}^{(0)}C_b)(\sigma^*)}\nn\\
   &:=&\tfrac{2\alpha_A}{\kappa} g_{AC}D_B p^{BC}(Y) -\tfrac{\alpha_A}{\alpha_0}T^0_{,A}\Pi_0(Y) \q
\ea
with $\cy^B(\sigma^*)=Y^B$. Here $D_B$ is the covariant derivative in $Y^B$--direction associated to the metric $g_{CD}$. 

As explained in section \ref{AbelianConstraints} multiplying the constraints (\ref{diff15}) by $g^{-\frac{1}{2}}(Y)$ gives diffeomorphism invariant Hamiltonian constraints that are Abelian.

The weakly Abelianized constraints are defined by 
\ba\label{diff17}
\tilde C(Y) &=& B^{-1} C_\perp(Y)  \q \text{with}\q\nn\\
B(Y)&=&   \{T^0(Y), C_\perp(Y)\}=g^{-\frac{1}{2}}(Y)(\Pi_0(Y)-\sum_{A=1}^3 T^0_{,A} \Pi_A(Y)) \q    .
\ea

%\ba \label{diff18} 
%T^0_{,A}(Y):=\frac{\partial}{\partial Y^A}T^0(Y)=D^\cy_{T^0,a \, {S^a}_A(\sigma^*)}
%\ea
%with $Y^B=\cy^B(\sigma^*),B=1,2,3$.

Note that the Hamiltonian constraints (\ref{diff15}) depend explicitly on the (physical) coordinates $\{Y^A\}_{A=1}^3$ via the potentials $V_{(K)}$. In particular (\ref{diff15}) does not transform as a density under transformations in $Y$. This is caused by the fact that the $\{Y^A\}_{A=1}^3$ are coordinates defined by the values of the physical fields $T^A$ and that these fields are coupled to the other degrees of freedom. An invariance of (\ref{diff15}) under transformations in $Y^A$ will only occur if the original Hamiltanion constraints are invariant under the corresponding transformations in the fields $T^A$. 
%backreaction of coordinates

%This metric depends on the spatial metric $g_{ab}$, on the matter fields $T^K$ and on the conjugated momenta $\Pi_K$. 
The  partially complete observable associated to the inverse metric $\gamma^{KL}$ (\ref{diff19}) also depends on the gravitational momentum $p^{AB}(Y)$ via the momenta $\Pi_A(Y)$:
\ba\label{diff20}
\gamma^{00}(Y)&=&T^0_{,A}T^0_{,B}g^{AB}(Y)-g^{-1} \Pi_0\Pi_0(Y) \nn \\
\gamma^{A0}(Y)&=&T^0_{,B}g^{AB}(Y)-g^{-1} \Pi_A\Pi_0(Y) \nn \\
\gamma^{AB}(Y)&=&g^{AB}(Y)-g^{-1} \Pi_A\Pi_B(Y)  \q  .
\ea

Finally the complete observable associated to $\gamma^{KL}$ is given by the formal power series
\ba
F_{[\gamma^{KL}(Y^*);T^0]}(\tau^0,\cdot)
=\sum_{r=0}^\infty \frac{1}{r!}\{\gamma^{KL}(Y^*),\tilde C[1]\}_r\,\, (\tau^0(Y^*)-T^0(Y^*))^r  \q .
\ea

\section{Summary}\label{summary}

We applied the concepts of partial and complete observables to general relativity. One main result is that one can reduce the number of constraints one has to deal with from infinity to one.

The two main ideas thereby were the following: firstly to use as partial observables phase space functions which are already invariant under almost all constraints. Such phase space functions are provided by canonical fields that behave as space--time scalars. Secondly for the calculation of complete observables we use weakly Abelian constraints. Under the application of these constraints `almost gauge invariant' phase space functions remain `almost invariant'. 
We hope that for the cases where one has to deal with just one constraint it is much simpler to develop an approximation scheme for complete observables. Here it may be possible to use already existing methods.

Also we showed how one can solve the diffeomorphism constraint and introduce spatial diffeomorphism invariant Hamiltonian constraints, which moreover can be made to be (strongly) Abelian. These Abelian diffeomorphism invariant Hamiltonian constraints can in principle be quantized on the Hilbert state of diffeomorphism invariant states, see \cite{master} for a related proposal. This is in contrast to the original Hamiltonian constraints, which are not diffeomorphism invariant and therefore cannot be promoted to operators on this Hilbert state. In Loop Quantum Gravity one therefore quantizes the Hamiltonian constraints on the kinematical Hilbert state (of states which are not diffeomorphism invariant), however the regularization procedure involves the topology of the diffeomorphism invariant Hilbert space in an intricate way, see \cite{tt}. The diffeomorphism invariant Hamiltonian constraints proposed in this work have the advantage that their structure is similar to the original constraints, hence it could be possible to use for their quantization techniques from \cite{tt}. However one has either to find clock variables which provide good spatial coordinates or propose methods how to deal with clock variables which do not satisfy the requirements from section \ref{solvediffeo} everywhere on $\cc_d$.

Furthermore we connected the notion of gauge invariant observables in the space--time picture, i.e. (space--time) quantities invariant under space--time diffeomorphism, with the notion of Dirac observables in the canonical picture, i.e. phase space functions which are invariant under the flow of the spatial diffeomorphism and Hamiltonian constraints. This might facilitate a comparison of covariant and canonical quantization procedures based on observables.  

An important role is played by the clock variables. We know that complete observables associated to arbitrary (bounded) phase space functions will exist and be globally well--defined, if the clock variables provide a good parametrization of the gauge orbits
%\footnote{i.e. for two arbitrary points $y \neq y'$ in the same gauge orbit we should have $T_k(x)\neq T_k(y)$ for at least one of the clock variables $\{T_j\}$}
. 
However also if the clock variables do not provide a good parametrization it may be possible to calculate complete observables, as is shown in \cite{bd1}. So one should choose the clock variables such that a parametrization is at least provided locally in phase space.  
     
We will end with a remark on the quantization of complete observables. These complete observables depend on the classical parameters $\tau$, which in some sense replace the classical time parameter $t$, that we use in non--relativistic quantum mechanics. A quantum complete observable will encode the probability to find outcomes of a certain measurement. This measurement is among other things characterized by the classical parameters $\tau$. These parameters could for instance prescribe values of matter fields, specifying the space--time region at which the measurement has to take place. However it is interesting to note that although the matter fields are quantized the parameters $\tau$ are just classical parameters.

\vspace{1cm}

{\Large \bf Acknowledgements }

It is a pleasure to thank Thomas Thiemann for lots of discussion and especially for asking about the Poisson brackets between the diffeomorphism invariant Hamiltonian constraints. This work was supported by a grant from the German National Merit Foundation.

\begin{appendix}

\section{Reconstruction of Space--Time Tensors}\label{appendix1}

The Reconstruction Theorem of Kucha\v{r} \cite{hyperspace} does not only state conditions for canonical fields to be reconstructable as space--time scalars. It gives also conditions for the normal and tangential components of higher order tensor fields. We will review these conditions and then show in more generality than in section \ref{spacetimemetric} how to express these fields in a $T^K$--frame and that the so obtained components behave as space--time scalars, i.e. have ultra--local Poisson brackets with the constraints.

To begin with consider a space--time covector $t_\mu$. Using formulas (\ref{inf3}) and (\ref{inf6}) we define the tangential and normal components of this covector by
\ba\label{ajuly1}
t_a(z):={Z^\mu_a}(\sigma)_{|z^\nu=Z^\nu(\sigma)} t_\mu(z) \q \text{and}\q t_\perp(z):=-n^\mu(z) t_\mu(z)  \q 
\ea
where $Z:\Sigma \rightarrow \cs$ is an embedding of $\Sigma$ into the space--time manifold $\cs$. The space--time covector can then be written as $t_\mu=t_a Z^a_\mu +t_\perp n_\mu$. 
The components $t_a$ and $t_\perp$ can be either seen as fields on the embedded hypersurface $Z(\Sigma)$ (which will however depend on the embedding $Z$, see below) or as fields on $\Sigma$: $t_j(\sigma)=t_j(z)_{|z^\nu=Z^\nu(\sigma)}$.

Now consider an embedding $Z_\varepsilon$ which maps the point $\sigma^*$ to the same space--time point as $Z$, i.e. which tilts the new embedding with respect to the old one:
\ba\label{ajuly2}
Z^\mu(\varepsilon,\sigma)=Z^\mu(0,\sigma)+\varepsilon \Lambda(\sigma)n^\mu(z)_{|z^\nu=Z^\nu(0,\sigma)}+O(\varepsilon^2)
\ea
where $Z^\mu(0,\sigma)\equiv Z^\mu(\sigma)$ and $\Lambda(\sigma^*)=0$. The new tangential basis $Z^\mu_{,a}(\varepsilon,\sigma^*)_{|(z^*)^\nu=Z^\nu(0,\sigma^*)}$ to the hypersurface $Z(\varepsilon,\Sigma)$ at the point $z^*=Z(0,\sigma^*)$ differs from the old one by
\ba\label{ajuly3}
Z^\mu_{,a}(\varepsilon,\sigma^*)_{|(z^*)^\nu=Z^\nu(0,\sigma^*)}=Z^\mu_{,a}(0,\sigma^*)+\varepsilon \Lambda_{,a}(\sigma^*)n^\mu(z^*) +O(\varepsilon^2)\q .
\ea
The new normal vector ${}^{(\varepsilon)}n^\mu$ can be found by using that $0={}^{(\varepsilon)}n^\mu \gamma_{\mu\nu} Z^\nu_{,a}(\varepsilon)$. Differentiating this equation with respect to $\varepsilon$ we will get
\ba\label{ajuly4}
Z^\mu_{,a}(0,\sigma^*)\gamma_{\mu\nu}\frac{\partial}{\partial\varepsilon}{}^{(\varepsilon)}n^\nu  \,(z^*)=\Lambda_{,a}  (\sigma^*) \q .
\ea
Multiplying both sides of this equation with $Z^a_{\mu'}(0,\sigma^*)$ and using $Z^a_{\mu'}Z^\mu_a=\delta^\mu_{\mu'}-n_{\mu'}n^\mu$ as well as $n_\nu \frac{\partial}{\partial \varepsilon}{}^{(\varepsilon)}n^\nu=0$ we arrive at
\ba\label{ajuly5}
{}^{(\varepsilon)}n^\mu(z^*)={}^{(0)}n^\mu(z^*)+\gamma^{\mu\nu}(z^*) Z^a_\mu(0,\sigma^*) \Lambda_{,a}(\sigma^*)+O(\varepsilon^2) \q .
\ea

Now the space--time covector $t_\mu$ at the point $z^*$ does of course not depend on the shape of the embedded hypersurface. The projections in (\ref{ajuly1}) therefore change to
\ba\label{ajuly6}
{}^{(\varepsilon)}t_a(z^*)&=&Z^\mu_a(\varepsilon,\sigma^*)t_\mu(z^*)\q \;={}^{(0)}t_a(z^*)-\varepsilon \Lambda_{,a}(\sigma^*){}^{(0)} t_\perp(z^*)+O(\varepsilon^2)  
\nn\\
{}^{(\varepsilon)}t_\perp(z^*)&=&(-1)\,{}^{(\varepsilon)}n^\mu \,t_\mu(z^*)\q={}^{(0)}t_\perp(z^*)-\varepsilon g^{ab}\Lambda_{,a}(\sigma^*){}^{(0)}t_b(z^*) +O(\varepsilon^2) \q .
\ea

On the other hand we know that the Hamiltonian constraint 
\ba\label{ajuly7}
C_\perp[\Lambda]:=\int_\Sigma \Lambda(\sigma) C_\perp(\sigma) \td^d\sigma
\ea
generates the deformation (\ref{ajuly2}) of the embedding of the spatial hypersurface $\Sigma$. A necessary condition for canonical fields $t_a(\sigma)$ and $t_\perp(\sigma)$ to be reconstructable as a space--time covector is therefore that
\ba\label{ajuly8}
\{t_a(\sigma), C_\perp[\Lambda]\} &=& -\Lambda_{,a}t_\perp (\sigma) \nn \\
\{t_\perp (\sigma), C_\perp[\Lambda]\} &=& -g^{ab}\Lambda_{,a}t_b (\sigma)
\ea
holds at least on the constraint hypersurface for all smearing functions $\Lambda$ with $\Lambda(\sigma)=0$.

A further requirement, that ensures the appropriate behaviour under diffeomorphisms tangential to the hypersurface is, that $t_a$ behaves as a spatial covector under spatial diffeomorphisms and $t_\perp$ as a spatial scalar, i.e.
\ba\label{ajuly9}
\{ t_a(\sigma),\vec{C}[\vec{\Lambda}]\}&=&\Lambda^b t_{a,b}(\sigma)+ \Lambda^b_{,a}t_b(\sigma)  \nn\\
\{t_\perp(\sigma),\vec{C}[\vec{\Lambda}]\}&=&\Lambda^b t_{\perp,b}(\sigma)  \q 
\ea 
for all smearing functions $\Lambda^b,b=1,\ldots,d$ where
\ba\label{ajuly10}
\vec{C}[\vec{\Lambda}]:=\int_\Sigma \Lambda^b(\sigma) C_b(\sigma) \td^d\sigma  \q .
\ea

%%%%%%%%%%%%%%%%%%%%%%
%%%%%%%%%%%%%%%%%%%%%
%%%%%%%%%%%%%%%%%%%%%%%%
%%%%%%%%%%%%%%%%%%%%%%%%

In \cite{hyperspace} it is shown that the conditions (\ref{ajuly8}) and (\ref{ajuly9}) are also sufficient in order that 
\ba\label{appe1}
t_\mu(z^*)={}^{(\varepsilon)}t_a(\sigma^*) Z^a_\mu(\varepsilon,\sigma^*)+{}^{(\varepsilon)}t_\perp(\sigma^*)\,{}^{(\varepsilon)}n_\mu(z^*)
\ea
does not depend on the embedding (\ref{ajuly2}) i.e. on the parameter $\varepsilon$. To this end one has to check the dependence of $Z^a_\mu(\varepsilon,\sigma^*)=\gamma_{\nu\mu}(z^*)g^{ab}(\sigma^*)Z_{b}^{\nu}(\varepsilon,\sigma^*)$ and ${}^{(\varepsilon)}n_\mu(z^*)=\gamma_{\mu\nu}(z^*){}^{(\varepsilon)}n^\nu(z^*)$ on $\varepsilon$. This dependence follows from the equations (\ref{ajuly3},\,\ref{ajuly5}) and the fact that neither $\gamma_{\mu\nu}(z^*)$ nor the inverse spatial metric $g^{ab}(\sigma^*)$ depends on $\varepsilon$. The independence of $g^{ab}(\sigma^*)$ can be seen by considering the spatial metric $g_{ab}$:
\ba\label{appe2}
g_{ab}(\sigma^*)&=&Z^\mu_{a}Z^\nu_{b}(\varepsilon,\sigma^*)\gamma_{\mu\nu}(z^*)
\nn\\
&=&
(Z^\mu_{a}(0,\sigma^*)+\varepsilon \Lambda_{,a}(\sigma^*)n^\mu(z^*))(Z^\nu_{b}(0,\sigma^*)+\varepsilon \Lambda_{,b}(\sigma^*)n^\nu(z^*))\gamma_{\mu\nu}(z^*) +O(\varepsilon^2)
\nn\\
&=&
Z^\mu_{a}Z^\nu_{b}(0,\sigma^*)\gamma_{\mu\nu}(z^*)+O(\varepsilon^2) \q .
\ea
Now using equations (\ref{ajuly6}) and (\ref{ajuly3},\,\ref{ajuly5}) in equation (\ref{appe1}) one can see that $t_\mu(z^*)$ does not depend on $\varepsilon$.
%%%%%%%%%%%%%%%%%%
%%%%%%%%%%%%%%%%%%%
%%%%%%%%%%%%%%%%
%%%%%%%%%%%%%%%%

Similarly to conditions for space--time covectors one can derive conditions for the tangential and normal components of higher order covaraint tensors. For instance the tangential, normal and mixed components of a second order covariant tensor are defined by contracting the two space--time indices either with $Z^\mu_{,a}$ and/or with $n^\mu$. As explained above these change under a tilt of the embedding and therefore the components have to show the following behaviour under the application of a smeared Hamiltonian constraint $C_\perp[\Lambda]$ with $\Lambda(\sigma)=0$:
\ba\label{ajuly11}
\{t_{ab}(\sigma), C_\perp[\Lambda]\}&=& -\Lambda_{,a}t_{\perp b}(\sigma)-\Lambda_{,b}t_{a \perp}(\sigma)\nn\\
\{t_{a\perp}(\sigma), C_\perp[\Lambda]\}&=& -\Lambda_{,a}t_{\perp \perp}(\sigma)-g^{bc}\Lambda_{,b}t_{a c}(\sigma)\nn\\
\{t_{\perp b}(\sigma), C_\perp[\Lambda]\}&=&-g^{ac}\Lambda_{,a}t_{cb}(\sigma)-\Lambda_{,b}t_{\perp \perp}(\sigma)\nn\\
\{t_{\perp\perp}(\sigma), C_\perp[\Lambda]\}&=& -g^{ab}\Lambda_{,a}t_{b\perp}(\sigma)  -g^{ab}\Lambda_{,a}t_{\perp b}(\sigma) \q .
\ea

Furthermore the components should have the appropriate behaviour under spatial diffeomorphisms:
\ba\label{ajuly12}
\{t_{ab}(\sigma),\vec{C}[\vec{\Lambda}]\}&=&
\Lambda^c t_{ab,c}(\sigma)+ \Lambda^c_{,a}t_{cb}(\sigma)+ \Lambda^c_{,b}t_{ac}(\sigma) 
\nn\\
\{t_{a\perp}(\sigma),\vec{C}[\vec{\Lambda}]\}&=&
\Lambda^c t_{a\perp,c}(\sigma)+ \Lambda^c_{,a}t_{c\perp}(\sigma) 
\nn\\
\{t_{\perp b}(\sigma),\vec{C}[\vec{\Lambda}]\}&=&
\Lambda^c t_{\perp b,c}(\sigma)+ \Lambda^c_{,b}t_{\perp c}(\sigma) 
\nn\\
\{t_{\perp \perp}(\sigma),\vec{C}[\vec{\Lambda}]\}&=&\Lambda^c t_{\perp \perp,c}(\sigma) \q .
\ea

In order to derive conditions for the components 
\ba\label{ajuly13}
r^a:=Z^a_\mu r^\mu \q\q \text{and}\q\q r^\perp:=n_\mu r^\mu
\ea
of a contravariant space--time vector consider the contraction of a space--time vector $r^\mu$ with a space--time covector $t_\mu$:
\ba
r^\mu t_\mu=(r^a Z^\mu_a+r^\perp n^\mu)(t_b Z^b_\mu+t_\perp n_\mu)= r^at_a-r^\perp t_\perp \q .
\ea
%The minus sign appears because of the normal--normal component of the Lronecker symbol: $n^\mu n_\nu\delta_\mu^\nu=-1$.
The expression $(r^at_a-r^\perp t_\perp)(\sigma)$ has to behave as a space--time scalar, i.e. the Poissin bracket with the Hamiltonian constraint $C_\perp[\Lambda]$ where $\Lambda(\sigma)=0$ has to vanish (at least on the constraint hypersurface). From that we arrive at the following conditions for the tangential and normal components of a contravariant space--time vector:
\ba\label{ajuly14}
\{r^a(\sigma), C_\perp[\Lambda]\}=-\Lambda_{,b}g^{ab} r^\perp (\sigma) \nn\\
\{r^\perp(\sigma), C_\perp[\Lambda]\}=-\Lambda_{,b}r^b (\sigma) \q .
\ea
The Poisson brackets with the diffeomorphism constraints have to be as follows:
\ba\label{ajuly15}
\{r^a(\sigma),\vec{C}[\vec{\Lambda}]\}&=& \Lambda^c {r^a}_{,c}(\sigma)-{\Lambda^a}_{,c}r^c(\sigma) \nn\\
\{r^\perp(\sigma),\vec{C}[\vec{\Lambda}]\}&=& \Lambda^c {r^\perp}_{,c}(\sigma) \q .
\ea
From here it is straightforward to give conditions for higher order contravariant tensors and mixed tensors.

To express a tensor field in $T^K$ components note that according to the equations (\ref{fields29},\,\ref{fields31}) $T^K_{,a}$ and $T^K_{\perp}$ behave as the tangential and normal components of a covariant space--time vector. Therfore the $T^K$ component of a contravariant space--time vector $r^\mu$ is given by
\ba\label{ajuly16}
r^K=T^K_{,\mu}r^\mu=T^K_a r^a -T^K_{,\perp}r^\perp \q .
\ea
That is, if we have canonical fields $r^a$ and $r^\perp$ satisfying the conditions (\ref{ajuly14}) and (\ref{ajuly15}) we know that $r^K:=T^K_a r^a -T^K_{,\perp}r^\perp$ are space--time scalars, i.e. have ultra--local Poisson brackets with the constraints $(C_\perp(\sigma),C_a(\sigma))$.

Similarly one can show that ${(B^{-1})^a}_K$ and ${(B^{-1})^\perp}_K$ satisfy the conditions for the tangential and normal components of a contravariant space--time vector. To this end consider the Poisson bracket of ${(B^{-1})^j}_K;\,j=\perp,a$ with a smeared Hamiltonian constraint
\ba\label{ajuly17}
\{{(B^{-1})^j}_K (\sigma), C_\perp[\Lambda]\}&=&-{(B^{-1})^j}_L{(B^{-1})^k}_K(\sigma) \{{B^L}_k (\sigma), C_\perp[\Lambda]\}  
\nn\\
&=&-{(B^{-1})^j}_L\bigg( {(B^{-1})^a}_K ( \Lambda {B^L}_\perp)_{,a}(\sigma)+ 
\nn\\ 
&&\q\q{(B^{-1})^\perp}_K ( \Lambda(\sigma) \{{B^L}_\perp(\sigma) , C_\perp[1]\} + \Lambda_{,a} g^{ab} {B^L}_b(\sigma))\bigg)
\nn\\
&=&\Lambda(\sigma) \{{(B^{-1})^j}_K (\sigma), C_\perp[1]\} 
\nn\\
&&\q  -\delta^j_\perp \Lambda_{,a}{(B^{-1})^a}_K (\sigma)
-\delta^j_b g^{ab} \Lambda_{,a} {(B^{-1})^\perp}_K(\sigma) \q .
\ea
Here we used that ${B^L}_a=T^L_{,a}$ and ${B^L}_\perp=-T^L_{,\perp}$ as well as the Poisson brackets (\ref{fields29}) and (\ref{fields31}).
According to equation (\ref{ajuly17}) the components ${(B^{-1})^a}_K$ and ${(B^{-1})^\perp}_K$ fulfill the requirements (\ref{ajuly14}). The requirements (\ref{ajuly15}) can be checked similarly:
\ba\label{ajuly18a}
\{ {(B^{-1})^j}_K (\sigma),  \vec{C}[\vec{\Lambda}]    \}  
&=& 
-{(B^{-1})^j}_L\bigg( 
{(B^{-1})^a}_K (\Lambda^b {B^L}_b)_{,a}(\sigma)
+ {(B^{-1})^\perp}_K \Lambda^b ({B^L}_\perp)_{,b}(\sigma) 
\bigg)  
\nn \\
&=& -\delta^j_b \Lambda^b_{,a}  {(B^{-1})^a}_K (\sigma)  
-\Lambda^b {(B^{-1})^j}_L \left({(B^{-1})^a
}_K {B^L}_a  \right)_{,b}(\sigma)
\nn\\
&&  + \delta^j_a \Lambda^b \left( {(B^{-1})^a}_K \right)_{,b}(\sigma)
 - \Lambda^b {(B^{-1})^j}_L \left({(B^{-1})^\perp}_K {B^L}_\perp  \right)_{,b}(\sigma)
\nn\\
&&+ \delta^j_\perp \Lambda^b \left({(B^{-1})^\perp}_K  \right)_{,b}(\sigma) 
\nn\\
&=& \Lambda^b \left( {(B^{-1})^j}_K\right)_{,b}(\sigma)
-\delta^j_b \Lambda^b_{,a}{(B^{-1})^a}_K(\sigma) \q\q .
\ea
Hence ${(B^{-1})^a}_K$ and ${(B^{-1})^\perp}_K$ are the components of a contravariant space--time vector. That is, if we have canonical fields $t_a$ and $t_\perp$ behaving a the components of a covariant space--time vector then
\ba\label{ajuly18}
t_K:={(B^{-1})^a}_K t_a-{(B^{-1})^\perp}_K t_\perp
\ea
will behave as space--time scalars. With this definition we will also have that $r^K t_K=r^at_a-r^\perp t_\perp$ is equal to the contraction of $r^\mu$ with $t_\mu$, therefore $t_K$ are the components of $t_\mu$ in the $T^K$--coordinates. 

Higher order tensor fields can be expressed in the $T^K$--basis by using the prescriptions (\ref{ajuly16}) and (\ref{ajuly18}) in order to contract contravariant or covariant indices respectively. If all indices are contracted in this manner the resulting fields will behave as space--time scalars in the sense that they will have ultra--local Poisson brackets with the diffeomorphism and the Hamiltonian constraints.

Note that the definition (\ref{fields25a}) of the $T^K$--metric components is consistent with this description. To see this remember that the tangential and normal components of the space--time metric are given by 
\begin{xalignat}{2}\label{ajuly19}
&\gamma_{\perp\perp}=n^\mu n^\nu \gamma_{\mu\nu}=-1  && \gamma_{\perp b}=-n^\mu Z^\nu_b \gamma_{\mu \nu}=0 \nn\\
& \gamma_{a \perp}=-n^\mu Z^\nu_a \gamma_{\nu \mu}=0 &&\gamma_{a \perp}=Z_a^\mu Z^\nu_b \gamma_{\mu \nu}=g_{ab} \q\q .
\end{xalignat}
Hence we have $\gamma_{KL}=
{(B^{-1})^a}_K{(B^{-1})^b}_L \,g_{ab}- 
{(B^{-1})^\perp}_K{(B^{-1})^\perp}_L$.

Finally, with the help of the complete observables associated to a certain choice of clock variables $T^K(\sigma),\sigma\in\Sigma,K=0,\ldots,d$ one can show that canonical fields which behave as space--time scalars transform also as space--time scalars under diffeomorphisms. With this we mean the following. Consider a canonical field $\phi(\sigma)$ behaving as a space--time scalar. Starting with some initial conditions, ie. at a certain phase space point $x$ evolve this field according to the equations (\ref{fields6}) to a one--parameter field $\phi(s,\sigma)$ and choose an one--parameter family of embeddings $Z_s$ as in (\ref{fields7}). We will end up with a solution $\phi(z)$ in some coordinate system $\{z^\mu\}_{\mu=0}^d$. For simplicity we will assume that the whole space--time manifold $\cs$ can be covered with one patch of coordinates $\{z^\mu\}_{\mu=0}^d$ taking values in some subset $\ck\in\Rl^{d+1}$. 

Now one can do the same procedure with differing lapse and shift functions and with a differing family of embeddings $Z'_{s'}$ but starting at the same phase space point $x$. That is one will end up with a different solution $\phi'(z')$ in some other coordinate system $\{{z'}^\mu\}_{\mu=0}^d$ and we will assume that these coordinates take values in the same subset $\ck$ as the coordinates above. The question is whether there is a space--time diffeomorphism that transforms the field $\phi(z)$ into $\phi'(z')$. 

We will assume that also the clock variables $T^K(\sigma)$ behave as space--time scalars and moreover the following: If one performs the above procedure with the clock variables the function $T: \ck \ni z\mapsto \{T^K(z)\}_{K=0}^d\in T(\ck)$ is bijective. The same should hold for $T':\ck \ni z'\mapsto \{{T'}^K(z')\}_{K=0}^d\in T'(\ck)$ with $T'(\ck)=T(\ck)$. Hence we can define the inverse function $z': T(\ck)\ni t \mapsto z'(t) \in \ck$ by $T'(z'(t))=t$ and with the help of this inverse function the coordinate transformation $\ck\ni z\mapsto z'=z'(T(z))\in \ck$.  

This means that we should have
\ba\label{append1}
\phi(z)=\phi'(z'(T(z)))
\ea
if $\phi$ transforms as a space--time scalar. We will show equation (\ref{append1}) using complete observables. Because of their interpretation we know that
\ba\label{append2}
\phi(z)&=&F_{[\phi(\sigma^*);\, T]}(\tau;\,x) \nn\\
\phi'(z')&=&F_{[\phi(\sigma^*);\, T]}(\tau';\,x)
\ea
where the fields $\tau(\sigma),\tau'(\sigma)$ have to satisfy $\tau^K(\sigma^*)=T^K(z)$ and ${\tau'}^K={T'}^K(z')$ respectively. Now, we need just to use that ${T'}^K(z'(T(z)))=T^K(z)$ in the second equation of (\ref{append2}), showing that indeed
\ba
\phi'(z'(T(z)))&=&F_{[\phi(\sigma^*);\, T]}(\tau;\,x)=\phi(z)
\ea
with $\tau^K(\sigma^*)={T'}^K(z'(T(z)))=T^K(z)$.

\end{appendix}


\vspace{2cm}



\begin{thebibliography}{99}
\addcontentsline{toc}{chapter}{\numberline{} Bibliography}
\parskip -1pt


\bibitem{1.1} %u
C. Rovelli, ``Loop Quantum Gravity", Living Rev. Rel. {\bf 1} (1998) 1,
[gr-qc/9710008]\\
T. Thiemann,``Lectures on Loop Quantum Gravity'', in {\it Quantum Gravity: From Theory to Experimental Search Proceedings, Bad Honnef, Germany 2002}, eds D. Giulini, C. Kiefer, C. L\"ammerzahl, LNP 631 (Springer,  Berlin 2003), [gr-qc/0210094]\\
A. Ashtekar, J. Lewandowski, ``Background Independent Quantum Gravity:
A Status Report'', Class. Quant. Grav. {\bf 21} (2004) R53;
[gr-qc/0404018]\\
L. Smolin, ``An Invitation to Loop Quantum Gravity'', [hep-th/0408048]

\bibitem{7.2} %u
C. Rovelli, ``Quantum Gravity'' (Cambridge University Press,
Cambridge 2004)

\bibitem{7.3} %u
T. Thiemann, ``Modern Canonical Quantum General 
Relativity'', (Cambridge University Press, Cambridge 2005),
[gr-qc/0110034]





\bibitem{ADM} R. Arnowitt, S. Deser, C.W. Misner, ``The Dynamics of General Relativity'' in {\it Gravitation: An Introduction to Current Research}, ed. by L. Witten (Wiley, New York, 1962), [gr-qc/0405109]

\bibitem{jacobson} 
J.N. Goldberg, J. Lewandowski, C. Stornaiolo: 
``Degeneracy in Loop Variables'',
Comm. Math. Phys. {\bf 148}, (1992),377; \\
T. Jacobson, J.D. Romano: ``The Spin Holonomy Group in General Relativity'',
Commun. Math. Phys. {\bf 155}, (1993), 261




\bibitem{RovPartObs} 
C. Rovelli,  ``Quantum mechanics without time: A model'', Phys. Rev. {\bf D42}, (1990), 2638\\
C. Rovelli, ``Time in quantum gravity: An hypothesis'', Phys. Rev. {\bf D43}, (1991), 442\\
C. Rovelli, ``Is There Incompatibility Between the Ways Time is Treated in General Relativity and in Standard Quantum Mechanics'' in {\it Conceptional Problems in Quantum Gravity}, ed. by A. Ashtekar and J. Stachel (Birkh\"auser, Boston, 1991), %p. 126
\\
C. Rovelli,
``What is observable in classical and quantum gravity?'',
Class. Quant. Grav. {\bf 8}, (1991) , 1895 \\
C. Rovelli, ``Partial Observables'',
Phys. Rev. {\bf D65}, (2002), 124013, [gr-qc/0110035] ;\\
C. Rovelli, ``Quantum Gravity'',
(Cambridge University Press, Cambridge 2004)



\bibitem{bd1} B. Dittrich, ``Partial and Complete Observables for Hamiltonian Constrained Systems'', [gr-qc/0411013]



\bibitem{reducedphasespaceq} T. Thiemann, ``Reduced Phase Space Quantization and Dirac Observables'', [gr-qc/0411031]

\bibitem{jb} J. Brunnemann, T. Thiemann, ``On (Cosmological) Singularity Avoidance in Loop Quantum Gravity'', [gr-qc/0505032]


\bibitem{wald} R. M. Wald, ``General Relativity'', (Chicago Univ. Press, 1984)

\bibitem{hyperspace} 
K.V. Kucha\v{r}, ``Geometry of Hyperspace'',
J. Math. Phys. {\bf 17}, (1977), 777\\
K.V. Kucha\v{r}, ``Kinematics of Tensor Fields in Hyperspace'',
J. Math. Phys. {\bf 17}, (1977) 792






\bibitem{BergmannHandbuch} 
P.G. Bergmann: ``General Theory of Relativity'' in {\it Encyclopedia of Physics, Vol. IV: Principles of Electrodynamics and Relativity}, ed. by S. Fl\"ugge (Springer, Berlin 1962)



\bibitem{deWitt} %B.S. DeWitt,  Phys. Rev {\bf 160}, (1967), 1113;\\
B.S. DeWitt, ``The quantization of geometry'', in {\it Gravitation: An Introduction to Current Research}, ed. by L. Witten (Wiley, New York 1962) 




\bibitem{dust} 
J.D. Brown, K.V. Kucha\v{r}, 
``Dust as a Standard of Space and Time in Canonical Quantum Gravity'',
Phys. Rev. {\bf D51} (1995), 5600, [gr-qc/9409001]





\bibitem{master} T. Thiemann, ``The Phoenix Project: Master Constraint 
Programme for Loop Quantum Gravity'', [gr-qc/0305080]\\
 B. Dittrich, T. Thiemann, ``Testing the Master Constraint
Programme for Loop Quantum Gravity I. General Framework'',
[gr-qc/0411138];
`` II. Finite Dimensional Systems'',
[gr-qc/0411139];
``III. SL(2,R) Models'',
[gr-qc/0411140];
`` IV. Free Field Theories'',
[gr-qc/0411141];
`` V. Interacting Field Theories'',
[gr-qc/0411142]




\bibitem{tt} %u
 T. Thiemann, ``Anomaly-free Formulation of non-perturbative,
four-dimensional Lorentzian Quantum Gravity", Physics Letters {\bf B380}
(1996) 257-264, [gr-qc/9606088];\\
T. Thiemann, ``Quantum Spin Dynamics (QSD)",
Class. Quantum Grav. {\bf 15} (1998) 839-73, [gr-qc/9606089];
``II. The Kernel of the Wheeler-DeWitt Constraint Operator",
Class. Quantum Grav. {\bf 15} (1998) 875-905, [gr-qc/9606090];
``III.
Quantum Constraint Algebra and Physical Scalar Product in Quantum General
Relativity", Class. Quantum Grav. {\bf 15} (1998) 1207-1247,
[gr-qc/9705017];
``IV. 2+1 Euclidean Quantum Gravity as a model to test 3+1
Lorentzian Quantum Gravity", Class. Quantum Grav. {\bf 15} (1998) 
1249-1280, [gr-qc/9705018]; 
``V. Quantum Gravity as the Natural Regulator of the Hamiltonian 
Constraint
of Matter Quantum Field Theories",
Class. Quantum Grav. {\bf 15} (1998) 1281-1314, [gr-qc/9705019];
``VI. Quantum Poincar\'e Algebra and a Quantum Positivity of Energy
Theorem for Canonical Quantum Gravity",
Class. Quantum Grav. {\bf 15} (1998) 1463-1485, [gr-qc/9705020]









\end{thebibliography}
\end{document}